\documentclass[12pt]{article}
\usepackage{amsmath}
\usepackage{amsfonts}        
\usepackage{amssymb}
\usepackage{amsbsy}
\usepackage{mathrsfs}
\usepackage{epsfig}      
\usepackage{color}       

\usepackage[T1]{fontenc} 

\usepackage{graphicx}

\newlength{\TZ}
\DeclareFontFamily{OT1}{pzc}{}
\DeclareFontShape{OT1}{pzc}{m}{it}{<-> s * [1.200] pzcmi7t}{}
\DeclareMathAlphabet{\mathpzc}{OT1}{pzc}{m}{it}
\newcommand{\BEQ}{\begin{equation}}     
\newcommand{\BEA}{\begin{eqnarray}}
\newcommand{\BD}{\begin{displaymath}}
\newcommand{\EEQ}{\end{equation}}       
\newcommand{\EEA}{\end{eqnarray}}
\newcommand{\ED}{\end{displaymath}}

\newcommand{\vep}{\varepsilon}          
\newcommand{\D}{{\rm d}}                
\newcommand{\II}{{\rm i}}               
\renewcommand{\Re}{{\rm Re\ }}          
\newcommand{\sgn}{{\rm sgn\,}}          
\newcommand{\demi}{\frac{1}{2}}         
\newcommand{\mel}[1]{\breve{#1}}        
\newcommand{\lap}[1]{\overline{#1}}     

\renewcommand{\vec}[1]{\boldsymbol{#1}} 
\newcommand{\fns}{\footnotesize}        

\newcommand{\appsection}[2]{\setcounter{equation}{0}\setcounter{subsection}{0}
\section*{Appendix #1. #2}
\renewcommand{\theequation}{#1.\arabic{equation}}
              \renewcommand{\thesection}{#1}
              \renewcommand{\thefigure}{#1\arabic{figure}}\setcounter{figure}{0} }

\catcode`\@=11
\def\numberbysection{\@addtoreset{equation}{section}
        \def\theequation{\thesection.\arabic{equation}}}
\numberbysection                        

\definecolor{gruen}{rgb}{0,0.625,0}       
\definecolor{rot}{rgb}{0.75,0,0}          
\definecolor{blau}{rgb}{0,0,0.75}         
\definecolor{casta}{rgb}{0.45,0.20,0}     
\definecolor{gelb}{rgb}{0.825,0.725,0.0}  

\begin{document}

\begin{titlepage}

\vskip 1.5 cm
\begin{center}
{\LARGE \bf Ageing in the exact correlations of the voter model on a fractal}
\end{center}

\vskip 2.0 cm
\centerline{{\bf Malte Henkel}$^{a,b}$}
\vskip 0.5 cm
\centerline{$^a$Laboratoire de Physique et Chimie Th\'eoriques (CNRS UMR 7019),}
\centerline{Universit\'e de Lorraine Nancy, B.P. 70239, F -- 54506 Vand{\oe}uvre-l\`es-Nancy Cedex, France}
\vspace{0.5cm}
\centerline{$^b$Centro de F\'{i}sica Te\'{o}rica e Computacional, Universidade de Lisboa,}
\centerline{Campo Grande, P -- 1749-016 Lisboa, Portugal}
\vspace{0.5cm}

\begin{abstract}
The exact behaviour of the enveloppes of the single-time and two-time correlators is found for the voter model on a fractal substrate, 
with nearest-neighbour interactions. Herein the geometry of the fractal substrate is described by its non-integer geometric fractal dimension $d_f$, and
its topology and diffusive transport by the distinct spectral dimension $d_s$. On the level of the equations of motion of the correlators this
can be modelled by considering a space-dependent diffusion constant ${\cal D}(r)\sim r^{-\theta}$ which implies the spectral index $\theta$, itself a function of $d_f$ and $d_s$. 
With a scaling ansatz, the generic phenomenology of ageing is confirmed and the dynamic exponent $\mathpzc{z}=2+\theta$ and the 
autocorrelation exponent $\lambda=d_f$ are derived. The explicitly found dynamic scaling functions are shown to depend only on the spectral dimension $d_s$. 
The decay of the enveloppe of the density of active interfaces with time is described by the exponent $\alpha=1-d_s/2$ for $d_s<2$, 
confirming the results of preexisting numerical simulations. 
\end{abstract}
\end{titlepage}

\setcounter{footnote}{0}

\section{Introduction: physical ageing and the voter model} \label{sec:1}

The understanding of the collective behaviour of many-body systems out of equilibrium continues to pose many challenges \cite{Taeu14,Cugl15,Giam16,Bait18,Bait22,Vinc24}.
{\em Physical ageing} \cite{Stru78} constitutes an often-studied example. It is defined by the properties \cite{Henk10}:
({\bf I}) slow relaxational dynamics, ({\bf II}) absence of time-translation-invariance and ({\bf III}) dynamical scaling.
Here, we shall be interested in classical systems, where ageing may be realised by preparing a system in a totally disordered initial state before quenching it instantaneously
either onto a critical point $T=T_c>0$ (called {\em non-equilibrium critical dynamics} \cite{Godr02}) or else into the disordered phase with temperature $T<T_c$
(called {\em phase-ordering kinetics} \cite{Bray94a} for a non-conserved order-parameter) although this generic phenomenology also applies to quantum systems. 
Microscopically, the system becomes inhomogeneous and will decompose into
clusters with a time-dependent linear size $\ell=\ell(t)$. If there is a late-time algebraic growth law $\ell(t)\sim t^{1/\mathpzc{z}}$, this defines
the {\em dynamic exponent} $\mathpzc{z}$. In situations where a continuum limit may be taken, 
conveniently one works with a coarse-grained time-space-dependent order-parameter $\phi(t,\vec{r})$
(i.e. the local magnetisation in magnetic systems). 
For a totally disordered initial state one usually admits that the average initial order-parameter vanishes, such that for all times
$\left\langle \phi(t,\vec{r})\right\rangle=\left\langle \phi(0,\vec{r})\right\rangle=0$.
The study of such systems is centred on analysing the {\em time-space correlation function} $C$ defined as
\BEQ \label{gl:1}
C(t,s;{r}) = \bigl\langle \phi(t,\vec{r})\phi(s,\vec{0})\bigr\rangle 
\EEQ
where the average is both over initial states as well as over thermal histories.
Since our study will use the continuum limit throughout, the habitual spatial translation- and rotation-invariances, such that $\vec{r}\mapsto r = |\vec{r}|$, 
will always be taken for granted, for the sake of notational and conceptual simplicity.
Setting $t=s$ in (\ref{gl:1}) gives\footnote{Practical means of obtaining the length scale $\ell(t)$ include
solving an equation $C\bigl(t;\ell(t)\bigr)=\mathfrak{c}$ with a constant $0<\mathfrak{c}<1$, or calculating the second moment
$\ell^2(t) = \left.\int_{\mathbb{R}^d}\!\D\vec{r}\: r^2\, C(t;\vec{r})\right/\int_{\mathbb{R}^d}\!\D\vec{r}\:C(t;\vec{r})$. 
They should all lead to the same long-time scaling $\ell(t)\sim t^{1/\mathpzc{z}}$.}
the {\em single-time correlator} $C(s;r) := C(s,s;r)$.
Setting $r=0$ in (\ref{gl:1}) produces the {\em two-time auto-correlator} $C(t,s):=C(t,s;{0})$. These will be the main object of our study.  
Throughout, we shall assume model-A-type dynamics without any macroscopic conservation law.

Sometimes, physical insight may be obtained from exactly solvable models, 
which at the least provide explicitly worked-out examples whose results may serve as input in more
generic conceptual studies. While most existing studies were confined to regular euclidean lattices, 
little is known about the ageing when the underlying lattice is fractal \cite{Mand83,Mand89}. 
This is of immediate relevance to systems which arise in lacunary matter, such as aerogels. 
Such an undertaking does require to clarify first how to characterise a fractal and second, 
which aspects of a lattice model may be brought to an exact analysis (see section~\ref{sec:2}). 
In this work, we shall study the physical ageing of the much-analysed {\em voter-model} \cite{Ligg85,Ligg99,Tome01,Krap10}, 
whose well-known exact solutions for $d\in\mathbb{N}$ were recently extended to any continuous $d\in\mathbb{R}_+$ \cite{Henk26}. 
Here, we shall exchange the euclidean lattice for a fractal lattice. 
Attention will be restricted to the {\em smooth enveloppes} \cite{Shau85a,Shau85b} 
of single-time and two-time correlators which can be obtained from the solution of certain differential equations. 
The additional log-periodic behaviour, due to the discrete scale-invariance \cite{Sorn98} 
of fractals and their irregularities at all length scales \cite{Bab08}, will not be considered here. 
We shall rather focus on the dependence of the enveloppes on (i) the non-integer geometric fractal dimension $d=d_f$, 
(ii) the spectral dimension $d_s$ or equivalently the spectral index $\theta=2 d_s/d-2$ which describes the non-trivial topology \cite{Alex82,Ramm83,Wilk84}. 
We shall mainly work with $d$ and $\theta$, in the voter model context considered as continuous parameters.

Formally, systems undergoing physical ageing obey the scaling form 
\begin{subequations} \label{gl:local}
\BEQ \label{gl:2}
C(t,s;{r}) = s^{-b} F_C\left( \frac{t}{s}; \frac{\bigl|\vec{r}\bigr|}{s^{1/\mathpzc{z}}}\right) 
\EEQ
with the {\em ageing exponent} $b$. For non-equilibrium critical dynamics at $T=T_c$ in pure magnetic systems, one generically expects that
$b=(d-2+\eta)/\mathpzc{z}$ \cite{Godr02,Cala05,Henk10,Taeu14}, where $\eta$ is a standard equilibrium critical exponent;
whereas for phase-ordering kinetics at $T<T_c$ in magnetic systems one expects $b=0$ and $\mathpzc{z}=2$, 
at least in pure systems with short-ranged interactions \cite{Bray94a,Henk10}. 
In addition, one usually finds asymptotically for $y\gg 1$  both
\BEQ \label{gl:3a}
f_C(y) = F_C(y;0) \sim y^{-\lambda/\mathpzc{z}} 
\EEQ
where $\lambda$ is the {\em auto-correlation exponent} \cite{Huse89}.  
\end{subequations}
The scaling function $F_{C}\bigl(y,r s^{-1/\mathpzc{z}}\bigr)$ in (\ref{gl:2}) is expected to be {\em universal},\footnote{The existence and asymptotic properties of
these scaling functions follow from a combination of dynamical scaling and generalised time-translation-invariance \cite{Henk25c}.} 
by which it is meant that its form should be independent of microscopic `details', such as the lattice structure,
the precise form of the interactions, etc. On the other hand, one expects it to be $d$-dependent (on a fractal, also $\theta$-dependent). 
A main objective in the study of exactly solvable systems is to derive exactly the values of the exponents $\mathpzc{z}, \lambda$ or $b$ and more generally to find the
scaling functions, for example $F_C(1;u)$ and $f_C(y)$. 

Another motivation for having undertaken this work comes from recent studies on tumor growth. Biological tissues in general are of a fractal nature and empirically, 
there exists evidence that the aggressivity of growing tumors increases with increasing fractal dimension $d$ \cite{Cros97,Elki22}. Analysis of simple `go-or-growth' models
of growing tumors suggests that in addition, the so-called spectral dimension $d_s$ may be relevant \cite{Fume25,Faja26}. We strive for an exactly solvable model where
the relative importance of both fractal dimensions can be explicitly worked out.

This work is organised as follows. 
In section~\ref{sec:2} we recall the definition of the nearest-neighbour voter model on an euclidean lattice and the characterisation of
fractals in terms of their characteristic parameters such as the fractal dimension $d$ and the spectral index $\theta$. 
We also briefly review some existing results on non-equilibrium exponents, notably for the critical contact process on a fractal. 
This is completed by a precise statement how the enveloppe of the single-time correlator can be obtained for the voter model from 
a diffusion equation with a space-dependent diffusion constant ${\cal D}(r)\sim r^{-\theta}$ \cite{Shau85a,Shau85b}. 
Section~\ref{sec:3} describes the exact calculations for the enveloppe of the single-time correlator, both for $d_s<2$ and $d_s>2$. 
The expected dynamical exponent $\mathpzc{z}=2+\theta$ is recovered. 
We also derive the time-dependent decay of the enveloppe of the density of active interfaces $n_{\rm r}(t)\sim t^{-\alpha}$, and predict the corresponding exponent $\alpha=1-d_s/2$ 
which for the first time explains the results of long-standing numerical simulations \cite{Such06,Bab08}. 
Section~\ref{sec:4} gives our derivation for the enveloppe of the two-time auto-correlator and yields $\lambda=d$, when $d_s<2$. 
However, all scaling functions, of the enveloppes of single-time and two-time correlators, only depend on the spectral dimension $d_s$. 
Section~\ref{sec:5} presents a discussion of our results.  
Four appendices present the technical details of our calculations. 

\section{The voter model} \label{sec:2}

\subsection{Voter model on regular euclidean lattices} 

The much-studied {\em voter model} is usually formulated as a classical spin model, defined in terms of spin variables $\sigma_{\vec{n}}=\pm 1$ attached to the sites
$\vec{n}\in\Lambda\subset\mathbb{Z}^d$ of a hyper-cubic lattice in $d$ dimensions. 
A spin configuration is denoted as $\{\sigma\} = \bigl(\sigma_1,\ldots,\sigma_{\cal N}\bigr)$,
where ${\cal N}=|\Lambda|$ is the total number of sites. That configuration arises with the probability $P\bigl(\{\sigma\};t\bigr)$.
The dynamics is described in terms of a master equation
\BEQ \label{gl:2.1}
\partial_t P\bigl( \{\sigma\};t) = \sum_{\{\sigma'\}} \left[ w\bigl(\{\sigma' \}\to \{\sigma \}\bigr) P\bigl(\{\sigma'\};t)
-w\bigl(\{\sigma \}\to \{\sigma' \}\bigr) P\bigl(\{\sigma\};t) \right]
\EEQ
In the voter model, transitions between configurations occur via single spin flips.
If the spin to be flipped is at site $\vec{n}$, the transition rates of the {\em voter model} are \cite{Ligg85,Ligg99,Tome01,Krap10}
\BEQ \label{gl:2.2}
w\bigl(\{\sigma \}\to \{\sigma' \}\bigr) ~~\mapsto~~
w_{\vec{n}}\bigl(\{\sigma \}\bigr) =  \demi\left( 1 - \frac{1}{2d} \sigma_{\vec{n}} \sum_{\vec{m}(\vec{n})} \sigma_{\vec{m}} \right)
\EEQ
where $\vec{m}(\vec{n})$ are the nearest-neighbour sites with respect to the site $\vec{n}\in\Lambda$. 
We shall specify shortly in what sense we can study this model on fractal. 
The voter model has apparently been first introduced in the 1960s in studies of genetic
correlations before receiving profound interest from probability theory, see \cite{Ligg85,Corb24e} and refs. therein.
Its numerous applications include: consideration as a model for opinion forming (possibly with generalisations) \cite{Cast09,Redn09,Fern14}, 
for modelling network dynamics \cite{Vasq08,Carr16,Doro22,Bern23}, 
as a non-equilibrium spin system subject to two distinct baths each creating its proper dynamics at its own temperature \cite{Droz89,Droz90}, 
as a model of surface catalytic reactions \cite{Krap92,Oliv03}
or finally as a prototype of critical-point ageing in models with several absorbing states \cite{Dorn01}. 
On any lattice, the voter model is dual to the diffusion-pair annihilation process, see \cite{Schu95}. 
It is one of the very few models of interacting spins which is analytically solvable in any number of dimensions $d\in\mathbb{R}_+$, 
for any kind of interactions \cite{Krap92,Frac96,Frac97,Vasq08,Corb24e,Corb24f,Henk26}.
With the widely studied Glauber-Ising model \cite{Glau63} it shares the invariance under a total
spin-reversal $\sigma_{\vec{n}}\mapsto -\sigma_{\vec{n}}$ for all $\vec{n}\in\Lambda$. In contrast to the latter one, the voter model 
for $d\ne 1$ does not satisfy the detailed-balance condition \cite{Ligg85,Ligg99,Tome01,Krap10,Godr13}; hence its stationary states cannot be at equilibrium. 
Rather, the voter model has two absorbing states,
namely $\sigma_{\vec{n}}=1$ for all $\vec{n}\in\Lambda$ (or $\sigma_{\vec{n}}=-1$ for all $\vec{n}$) into which the system
may enter but it cannot leave. For long-ranged interactions with distant-dependent interaction rates
$w(r)\sim r^{-\alpha}$, its behaviour has been analysed in detail \cite{Rodr11,Tart15,Corb24a,Corb24b,Corb24c,Corb24d,Corb24e,Corb24f,Godr24} in $d=1,2,3$ dimensions. 
For $\alpha<2+d$, if dynamical scaling holds at all, the dynamical exponent $\mathpzc{z}<2$ \cite{Corb24a,Corb24b,Corb24e} which implies a super-diffusive behaviour
for the characteristic length scale. The dynamics of the density $n_{\rm r}(t)$ of reactive $`+-'$-interfaces gives another important characteristics. 
For the example of short-range interactions, the long-time behaviour is \cite{Frac96,Oliv03,Cast09}
\BEQ \label{gl:densact}
n_{\rm r}(t) = \demi \biggl( 1 - \bigl\langle \sigma_{\vec{n}} \sigma_{\vec{n}+\vec{1}} \bigr\rangle \biggr) \sim
\left\{ \begin{array}{ll} t^{-(2-d)/2}                         & \mbox{\rm ~~;~ if $d<2$} \\
                          1/\ln t                              & \mbox{\rm ~~;~ if $d=2$} \\
                          \mathfrak{a} - \mathfrak{b} t^{-d/2} & \mbox{\rm ~~;~ if $d>2$}
        \end{array} \right.
\EEQ
where $\mathfrak{a},\mathfrak{b}$ are constants. In the litt\'erature, this is referred to by the statement that for $d\leq 2$ the voter model undergoes {\em coarsening} since 
$n_{\rm r}(t)\to 0$. For $d>2$, since $n_{\rm r}(t)$ remains finite, there are infinitely many stationary states  \cite{Ligg85,Ligg99}. 

The resolubility of the voter model can be cast in the following form. For technical simplicity, it is convenient to use a spatial continuum limit such that
the single-time spin-spin correlator $C_{\vec{n}}(t)=\bigl\langle \sigma_{\vec{n}}(t) \sigma_{\vec{0}}(t)\bigr\rangle$ turns into the function 
$C(t;\vec{r})=\bigl\langle \phi(t,\vec{r})\phi(t,\vec{0})\bigr\rangle$ in terms of a coarse-grained order-parameter $\phi(t,\vec{r})$. 
Similarly, the two-time spin-correlator $C_{\vec{n}}(t,s)=\bigl\langle \sigma_{\vec{n}}(t) \sigma_{\vec{0}}(s)\bigr\rangle$ becomes the function  
$C(t,s;\vec{r})=\bigl\langle \phi(t,\vec{r})\phi(s,\vec{0})\bigr\rangle$.  
On a regular (hyper-cubic) lattice, eqs.~(\ref{gl:2.1},\ref{gl:2.2}) imply the equations of motion
\BEQ \label{gl:electeur}
\partial_t C(t;\vec{r}) = \Delta_{\vec{r}} C(t;\vec{r}) \;\; , \;\; \partial_t C(t,s;\vec{r}) = \demi \Delta_{\vec{r}} C(t,s;\vec{r}) \;\; ; \;\; C(s,s;\vec{r})=C(s;\vec{r}) 
\EEQ
where $\Delta_{\vec{r}}$ is the spatial laplacian and diffusion constants were scaled to unity. 
In principle, there is the constraint $C(t;\vec{0})=1$ and some initial condition $C(0;\vec{r})=C_0(\vec{r})$ should be specified as well. 
If one uses a scaling ansatz to simplify the calculations, the initial correlator $C_0(\vec{r})$ 
may become implicit \cite{Bray97,Corb24b,Corb24c,Corb24e,Henk26} and one may have to reconsider the applicability of the constraint 
\cite{Krap10,Corb24e,Henk26} (see also section~\ref{sec:3}). 

\subsection{Fractal substrates}

\begin{table}[tb]
\begin{center}
\begin{tabular}{|c|c|llll|c|}  \hline
name                       &                                                                            & ~$d_f$   & ~$d_w$      & ~$d_s$       & ~$\theta$  & Ref. \\ \hline
Sierpinksi triangle        & \includegraphics[height=0.037\hsize]{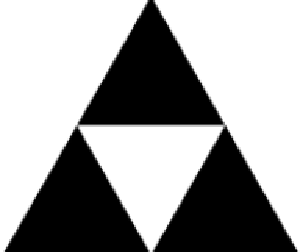} & $1.5850$ & $2.3219$    & $1.3652$     & $0.3219$   & \cite{Ramm83} \\ 
checkerboard lattice       & \includegraphics[height=0.037\hsize]{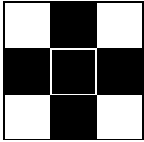}    & $1.4650$ & $2.4651$    & $1.1886$     & $0.4651$   & \cite{Chen96,Schu00} \\[0.1cm] 
Sierpinksi carpet SC(3,1)  & \includegraphics[height=0.037\hsize]{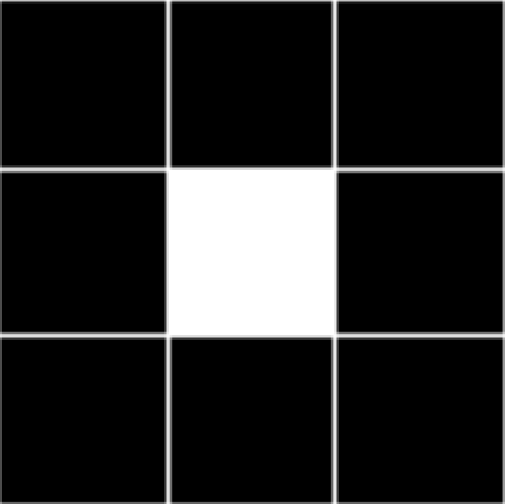}       & $1.893$  & $2.101$     & $1.802$      & $0.101$    & \cite{Song17}\\
                           &                                                                            & $1.893$  & {\fns $2.097$} & $1.80525$ & {\fns $0.097$} & \cite{Barl90} \\[0.15cm] 
Sierpinksi carpet SC(3,3)  & \includegraphics[height=0.037\hsize]{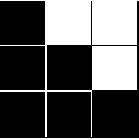}        & $1.6309$ & $2.5448$    & $1.2818$     & $0.5448$   & \cite{Schu00} \\[0.2cm] 
Sierpinksi carpet SC(4,6)  & \includegraphics[height=0.037\hsize]{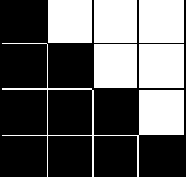}        & $1.6610$ & $2.5094$    & $1.3238$     & $0.5094$   & \cite{Schu00} \\
                           &                                                                            & $1.6610$ & $2.514(2)$  & {\fns $1.321(1)$} & $0.514(2)$ & \cite{Dasg99} \\[0.1cm] 
Sierpinksi carpet SC(5,10) & \includegraphics[height=0.037\hsize]{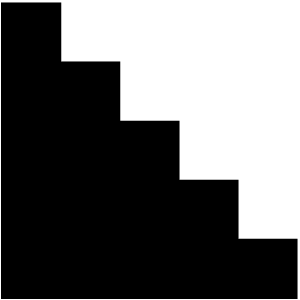}           & $1.6826$ & $2.4842$    & $1.3546$     & $0.4842$   & \cite{Schu00,Fran26} \\ \hline 
                           & $d=2$                                                                      & $1.9858$ & $2.8784(4)$ & $1.3100(11)$ & $0.8784(4)$ & \cite{Gras99} \\ 
percolation cluster        & $d=3$                                                                      & $2.52293(10)$ & {\fns $3.8$} & $1.32(6)$ & {\fns $1.8$} & \cite{Xu14,Argy84}\\ 
                           & $d>6$                                                                      & $4$      & {\fns $6$}  & $4/3$        & {\fns $4$} & \cite{Alex82} \\ \hline
silicia aerogel            & $3D$                                                                       & $2.364(20)$ &          & $1.252(61)$  &            & \cite{Cour87,Fume25} \\[-0.1cm] 
(experiment)               &                                                                            &  &          &   &            &  \\ \hline
\end{tabular}\end{center}
\caption[tab1]{Some examples of fractals (with the generating cell for deterministic fractals) and several characteristic fractal dimensions. 
The percolation cluster in $d$ spatial dimensions is studied at the critical point. \\
$d_f$: geometric fractal dimension, $d_w$: random walk dimension, $d_s$: spectral (fracton) dimension, $\theta$: spectral index. 
If necessary, one may use $d_w=2+\theta=\frac{2d_f}{d_s}$ (set in more small characters). 
}
\label{tab:1}
\end{table}

A {\em fractal} is a geometric object which is (statistically) similar at all scales \cite{Mand89,benA00}. 
Its properties are characterised by several dimensions which in general have non-integer, non-trivial
values. One of then is the {\em geometric fractal dimension} $d_f$, 
which describes the number of identical copies of the fractal under rescaling. Another is the {\em spectral dimension} $d_s$ 
(or fracton dimension \cite{Alex82,Ramm83}) which is related to the conductivity \cite{Berg24} 
on the fractal and carries information about the time-dependent topological properties of the fractal. 
Table~\ref{tab:1} gives for some deterministic fractals the defining generating cell from which by iteration the fractal can be constructed, 
by keeping the dark parts and suppressing the white ones. Then numerical values for 
the fractal dimensions $d_f$ and $d_s$ along with the spectral index $\theta=d_w-2$ are listed. We shall limit ourselves here to
examples which will be used in this work and complement this with an example of a stochastic fractal 
(critical percolation clusters\footnote{It is by now generally admitted that the Alexander-Orbach conjecture 
$d_s\stackrel{?}{=}\frac{4}{3}$ in percolation only holds true for $d>6$ and otherwise merely furnishes a good numerical approximation \cite{Naka94}.}) 
and an experiment on aerogel. It is remarkable how little $d_s$ changes in general between different fractals. 
Often the spectral dimension is not found directly, but rather the so-called walk dimension
\BEQ \label{gl:d_spectral}
d_w = 2+\theta = 2 \frac{d_f}{d_s}
\EEQ
as can be understood via scaling arguments on the energy density of states \cite{benA00}. 
For reviews, further background and much more long tables of numerical values we refer to \cite{Naka94,benA00,Schu00,Fran06,Song17,Bala20,Pati23}. 
It is well-known that the characterisation of a physical process on a fractal substrate
requires not only the geometric fractal dimension $d_f$ but also topological properties as encoded, e.g. in $d_s$. 
In what follows, we shall mainly use the spectral index $\theta$.

\begin{figure}[tb]
\includegraphics[width=0.3\hsize]{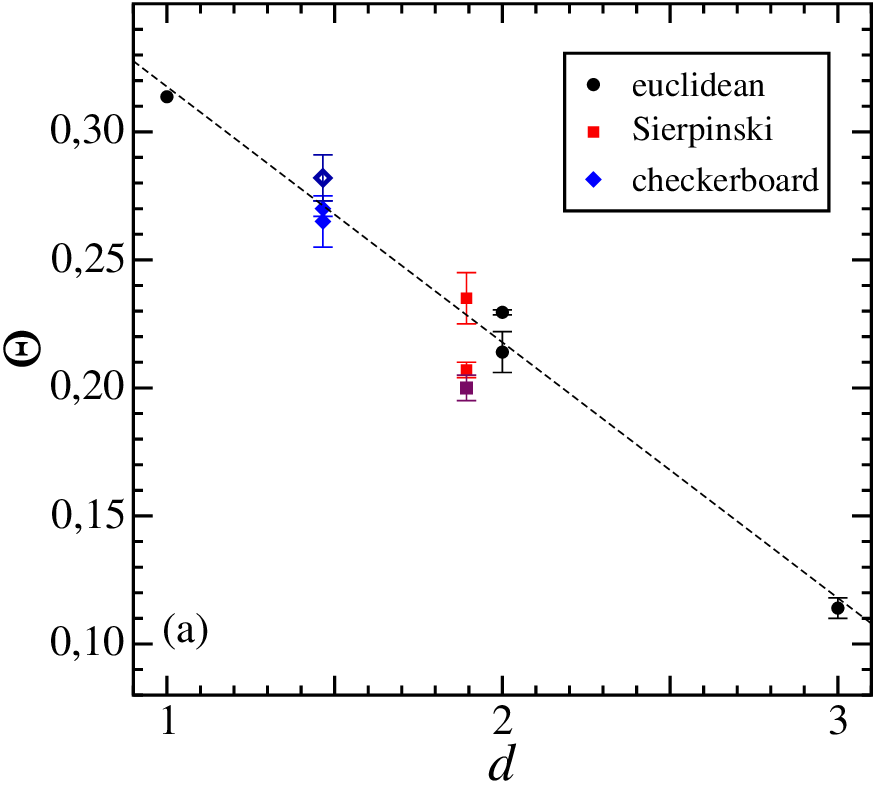}  ~
\includegraphics[width=0.3\hsize]{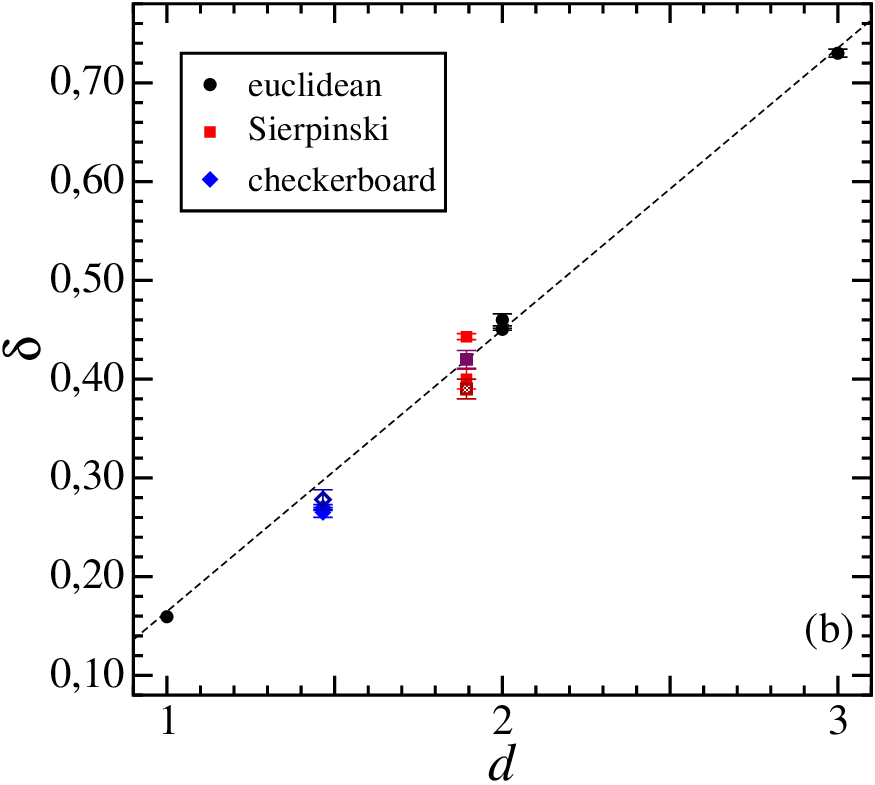}  ~
\includegraphics[width=0.3\hsize]{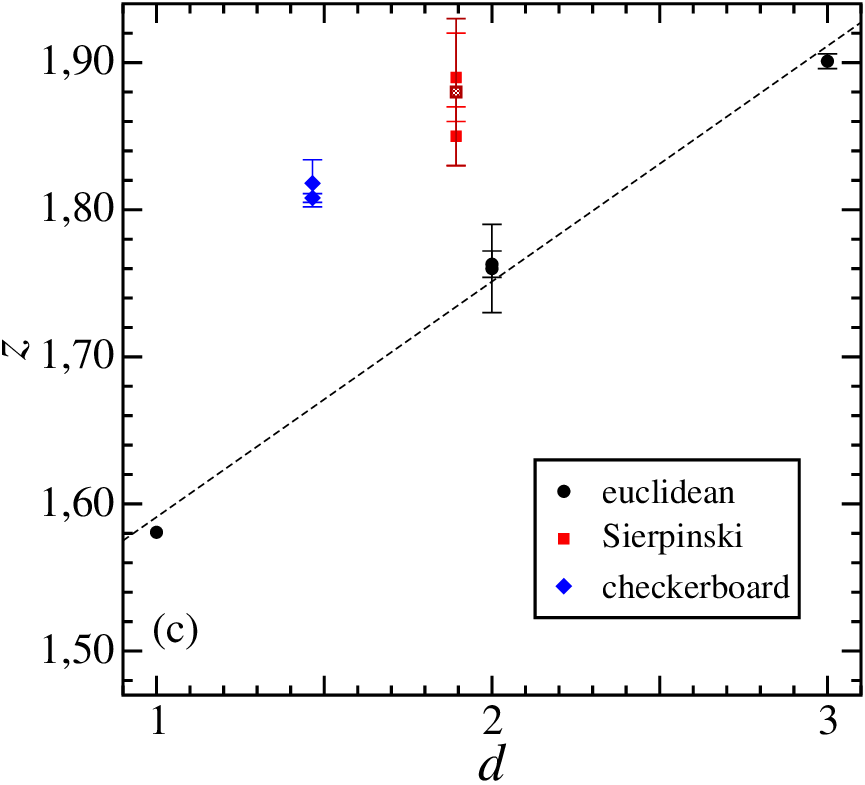}  ~
\caption[fig1]{Dependence of several non-equilibrium exponents of the critical contact process on the geometric fractal dimension $d=d_f$.
The exponents are: (a) the initial slip exponent $\Theta$ (b) the survival probability exponent $\delta$ (c) the dynamical exponent $\mathpzc{z}$. 
The black round dots correspond to the euclidean lattices and the linear dotted lines are meant as guides to the eye. 
The data points come from the checkerboard lattice and the Sierpinski carpet SC(3,1) \cite{Jens91,Lee08,Bab09,Argo15}. 
\label{fig1} }
\end{figure}

When considering the non-equilibrium evolution of physical systems on a fractal substrate, 
it must be bourne in mind that deterministic fractals merely obey a discrete scale-invariance \cite{Sorn98}. 
Beyond the usually expected dynamical scaling, this will lead to additional log-periodic oscillations \cite{Shau85a,Shau85b,Jens91,Naka94,Such06,Lee08,Bab08,Bab09,Argo15}.
Often, one rather studies the behaviour of the {\em enveloppe} of physical observables \cite{Shau85a,Shau85b} and then chooses to ignore the log-periodic oscillations which
in practise often have rather small amplitudes. 
Throughout this work, we shall focus exclusively on the behaviour of the enveloppes and do not consider any log-periodic oscillations around this. 
As an example on how the fractality of the substrate will modify the critical behaviour of many-body systems (at least via the fact that $d=d_f$ is not an integer), we consider
the critical contact process\footnote{We briefly recall its definition: on each site of a lattice, the system can assume one of two states ($X$, $Y$) 
such that $Y\stackrel{1-p}{\longrightarrow}X$ and between nearest neighbours the reaction 
$X+Y\stackrel{p}{\longrightarrow}2Y$ may occur, where $p$ is a control parameter. For $p<p_c$, the system falls back to an absorbing state whereas
for $p>p_c$ a finite density of the state $Y$ survives, e.g. \cite{Henk09}.} on fractal substrate. In figure~\ref{fig1}, we illustrate results on several exponents in the 
critical contact process, defined via the long-time asymptotics \cite{Henk09}
\BEQ
N(t) \sim t^{\Theta} \;\; , \;\; P(t) \sim t^{-\delta} \;\; , \;\; \ell(t) \sim t^{1/\mathpzc{z}}
\EEQ
where $N(t)$ is the average number of occupied sites,\footnote{The critical initial slip exponent $\Theta$ should not be confused with the fractal spectral index $\theta$.} 
$P(t)$ is the survival probability and $\ell(t)$ is a characteristic length scale of correlated clusters. 
Herein, we compare the values of these three exponents on euclidean lattices (taken from \cite{Henk09}), as a function  of the spatial dimension $d$, with existing numerical estimates 
\cite{Jens91,Lee08,Bab09,Argo15} on the fractal checkerboard and Sierpinski carpet lattices and $d=d_f$ (see table~\ref{tab:1}). 
The scattering in the data gives some {\it a posteriori} estimate on realistic error bars of these results. The data in figure~\ref{fig1}ab indicate
that the fractal lattices appear to interpolate quite well the values of the exponents $\Theta$, and especially $\delta$, in between the euclidean lattices with integer dimension.
Any further dependence, i.e. on the spectral dimension $d_s$, should be numerically small. 
However, figure~\ref{fig1}c shows that for the dynamical exponent $\mathpzc{z}$, this is very different.
The data from the fractal substrates dot not interpolate smoothly at all between the euclidean lattices.  
This strongly indicates an important dependence on other characteristics of the fractal, such as $d_s$ (or $\theta$). 
It is presently not understood why some exponents should only depend on the fractal geometry via $d_f$, while others
also appear to depend on other characteristics of the fractal substrate. 

This example illustrates that the behaviour of the enveloppes of physical observables on fractal substrates can be subtle. 
Our results in this work include the derivation of the exact behaviour of the enveloppes of single-time and two-time correlators in the voter model. 
In particular, for $d_s<2$, the density $n_{\rm r}(t) \sim t^{-\alpha}$ of reactive interfaces will be derived and generalises (\ref{gl:densact}). 
In the past, this was only compared with the result (\ref{gl:densact}) of euclidean lattices, viz. $\alpha_{\rm eucl}=1-d/2$,  which numerically is far off the mark. 
We shall show that on fractals, one rather has $\alpha=1-d_s/2$. Comparing our result with the findings of the available numerical studies \cite{Such06,Bab08}, 
we shall find a much better agreement, see section~\ref{sec:3}.2.

\subsection{The voter model on fractal lattices}

The relative ease by which a scaling approach produces the exact solution of the voter model for $d\in\mathbb{R}_+$ \cite{Henk26} makes it desirable to re-use
this technique for fractals as well. Since the equations of motion (\ref{gl:electeur}) of the euclidean voter model take the form of diffusion equations, it should be convenient
to use inspiration from the diffusion equation on fractals for a convenient generalisation in the voter model. 
In view of the discrete scale-invariance \cite{Sorn98}, the time-dependent probability $P(t,r)$ for a random walker to be at time $t$ in the shell
between radii $r$ and $r+\D r$ should be expected to be highly non-analytic and to display discontinuities at all length scales 
(and the known log-periodic oscillations are but one manifestation of this as exemplified in \cite{Such06,Bab08}). Therefore, one may only hope for progress
when restricting to the {\em enveloppes} of observables \cite{Shau85a,Shau85b} and approximate their description in terms of a differential equation. 
Herein, one tries to take into account the possibility of non-trivial topological effects through a spectral index \cite{Alex82,Ramm83,Wilk84} 
$\theta$ and an effective space-dependent diffusion constant ${\cal D}(r)\sim r^{-\theta}$ but
where spatial rotation-invariance is maintained. 
This analogy with diffusion on a fractal suggests as a starting point, for the calculation of the enveloppe of the single-time correlator \cite{Shau85a,Shau85b} 
\BEQ 
\partial_t C(t;r)  = \frac{1}{r^{d-1}} \frac{\partial}{\partial r} \left( r^{d-1-\theta} \,\frac{\partial C(t;r)}{\partial r} \right) 
\EEQ
where $d=d_f$ is the geometric fractal dimension.  
In doing so, we suppress any log-periodic oscillations. The resulting enveloppe will be a smooth function of its variables and one can study the influence the of the fractal
substrate, as encoded here in $d$ and $\theta$. Clearly, we expect to recover a length scaling $\langle r^2\rangle(t) \sim \ell^2(t) \sim t^{2/(2+\theta)}$. 
A similar generalisation will be given later for the enveloppe of the two-time correlator $C(t,s;r)$ and which will use the single-time correlator as initial condition. 

In what follows, it will always be understood that the time-space-dependent enveloppes will be studied, although we shall for brevity not mention it explicitly in general. 
Up to now, we have presented the necessary background and caveats and shall proceed with the derivation of the explicit solutions. 

\section{Single-time correlator} \label{sec:3}

Throughout this work, the spatial continuum limit will be taken. 
The enveloppe of the single-time spin-spin correlator $C_{\vec{n}}(t)=\bigl\langle \sigma_{\vec{n}} \sigma_{\vec{0}}\bigr\rangle(t)$ is cast as 
$C(t;\vec{r})=\bigl\langle \phi(t,\vec{r})\phi(t,\vec{0})\bigr\rangle$ in terms of a coarse-grained order-parameter $\phi(t,\vec{r})$ and in the voter model
obeys the diffusion equation $\partial_t C=\Delta_{\vec{r}}C$ as equation of motion. 
Since in the continuum limit, spatial translation- and rotation-invariance can be admitted, the enveloppe
of the correlator depends only on $r=|\vec{r}|$ and 
one writes on a fractal this equation in the form \cite{Shau85a,Shau85b,Fume25,Faja26} 
\BEQ \label{gl:2.4}
\partial_t C(t;r) = \Delta_{\vec{r}} C(t;r) = \frac{1}{r^{d-1}} \frac{\partial}{\partial r} \left( r^{d-1-\theta} \,\frac{\partial C(t;r)}{\partial r} \right) 
\EEQ
where $\Delta_{\vec{r}}$ is the spatial laplacian, $d=d_f$ is the geometric fractal dimension and $\theta=2 d/d_s -2$ is the spectral index 
(for some exemplary values, consult table~\ref{tab:1}). 
We shall always treat both $d>0$ and $\theta>0$ as continuous parameters.\footnote{Fractal Cantor tartans are examples with $d=d_s\not\in\mathbb{N}$ non-integer, but $\theta=0$ \cite{Bala18}.} 
Eq.~(\ref{gl:2.4}) will be the starting point for all subsequent 
calculations.\footnote{These calculations are made difficult by the discrete-lattice constraint $C_{\vec{0}}(t)=1$ for all $t\geq 0$, to be discussed in detail below.}

To solve this equation, we use the scaling ansatz \cite{Krap92,Tome01,Corb24e,Henk26} 
\BEQ \label{gl:2.5}
C(t;r) = t^{-b} f_C\bigl( \mathfrak{u} \bigr) \;\; , \;\; \mathfrak{u} = r\, t^{-1/(2+\theta)}
\EEQ
where $b$ will be identified later on with one of the ageing exponents. 
A separate time-dependence cancels if one admits $\mathpzc{z}=2+\theta>2$.\footnote{This is the opposite to the 
finding $\mathpzc{z}<2$ in long-range versions of the voter model \cite{Corb24a,Corb24b,Corb24e}.}
Eq.~(\ref{gl:2.5}) implies the double scaling limit $t\to\infty$, $r\to\infty$ such that $\mathfrak{u}$ is kept fixed.
It follows that the scaling function obeys the differential equation
\BEQ \label{gl:2.6}
f_C''(\mathfrak{u}) + \frac{d-1-\theta}{\mathfrak{u}}f_C'(\mathfrak{u}) +\frac{\mathfrak{u}^{1+\theta}}{2+\theta} f_C'(\mathfrak{u}) + b \mathfrak{u}^{\theta} f_C(\mathfrak{u}) = 0
\EEQ
Clearly, for large arguments $\mathfrak{u}\gg 1$, that is for physical distances much larger than the linear cluster size $\ell(t)\sim t^{1/(2+\theta)}$, spins should become
uncorrelated which leads to the first boundary condition $f_C(\mathfrak{u})\to 0$, when $\mathfrak{u}\to\infty$. 

To make further progress, we must find the value of the exponent $b$. This follows from the stationary solution which obeys \cite{Tome01}
\BEQ
\Delta_{\vec{r}} C(\infty;r) = 0 ~~\Longrightarrow~~ 
\frac{1}{\mathfrak{u}^{d-1}} \frac{\partial}{\partial\mathfrak{u}} \left( \mathfrak{u}^{d-1-\theta}\, \frac{\partial C}{\partial \mathfrak{u}} \right)  =0
\EEQ
with the promptly found solution $C(\mathfrak{u}) = \mathfrak{c}_0 + \mathfrak{c}_1 \mathfrak{u}^{2+\theta-d}$ where $\mathfrak{c}_{0,1}$ are constants. 
Because of the definition (\ref{gl:2.5}) of $\mathfrak{u}$, the stationary limit $t\to\infty$ corresponds to $\mathfrak{u}\ll 1$. 
In that limit, we have for $d<2+\theta$ that $C(\mathfrak{u})\to \mathfrak{c}_0$
saturates and for $d>2+\theta$ that $C(\mathfrak{u})\to \mathfrak{c}_1 \mathfrak{u}^{2+\theta-d}$ is reminiscent of an equilibrium  critical correlator
$C_{\rm eq}(r)\sim r^{-(d-2+\eta)}$. Combining this with the scaling from (\ref{gl:2.5}) allows us to conclude
\BEQ \label{gl:b-exp}
b = \left\{ \begin{array}{ll}  0                           & \mbox{\rm ~~;~ if $d<2+\theta$} \\
                               \frac{d-2-\theta}{2+\theta} & \mbox{\rm ~~;~ if $d>2+\theta$}
            \end{array} \right.
\EEQ
and for $\theta=0$ one reproduces the nearest-neighbour voter model result \cite{Henk26}. 
The case $d=2+\theta$ requires a separate treatment. 
Eq.~(\ref{gl:b-exp}) furnishes the second boundary condition on $f_C(\mathfrak{u})$, when $\mathfrak{u}\to 0$. 

\subsection{The case $d<2+\theta$} 
For $d<2+\theta$ (or $d_s<2$), we have $b=0$. 
Setting $g(\mathfrak{u})=f_C'(\mathfrak{u})$, we obtain the first-order differential equation
\BEQ
g'(\mathfrak{u}) + \left( d-1-\theta + \frac{\mathfrak{u}^{2+\theta}}{2+\theta} \right) \frac{1}{\mathfrak{u}} g(\mathfrak{u}) = 0
\EEQ
Then the solution proceeds via two standard integrations. In the first step, we have 
$g(\mathfrak{u})\sim u^{-d+1+\theta} e^{-\mathfrak{u}^{2+\theta}/(2+\theta)^2}$ and second, we find  
\BEQ
f_C(\mathfrak{u}) = F_0 + F_1\, \Gamma\left( 1 - \frac{d}{2+\theta}, \frac{\mathfrak{u}^{2+\theta}}{(2+\theta)^2} \right) \;\; , \;\; d<2+\theta
\EEQ
with the incomplete Gamma function $\Gamma(a,x)$ \cite{Abra65} and the constants $F_{0,1}$. 
Now, the first boundary condition $f_C(\infty)=0$ gives $F_0=0$. On the other  hand, if $\mathfrak{u}\ll 1$, 
the physical discrete-lattice constraint $C_{\vec{0}}(t)\stackrel{!}{=}1$ produces, because of $b=0$ from (\ref{gl:b-exp}), the normalisation $f_C(0)=1$. 
This fixes $F_1$ as well such that finally the time-space correlator is, in the scaling regime
\begin{subequations} \label{gl:voter-C1}   
\BEQ \label{gl:voter-C1-bas}               
C(t;r)  
= \frac{\Gamma\bigl(1-\frac{d}{2+\theta},\frac{1}{(2+\theta)^2}\frac{r^{2+\theta}}{t}\bigr)}{\Gamma\bigl(1-\frac{d}{2+\theta}\bigr)} \;\; ; \;\; d<2+\theta
\EEQ
\end{subequations}                        
It does agree with the generic scaling expectation (\ref{gl:2}). For $\theta=0$, eq.~(\ref{gl:voter-C1-bas}) reduces to the known exact result \cite{Henk26}. 

\begin{figure}[tb]
\includegraphics[width=0.3\hsize]{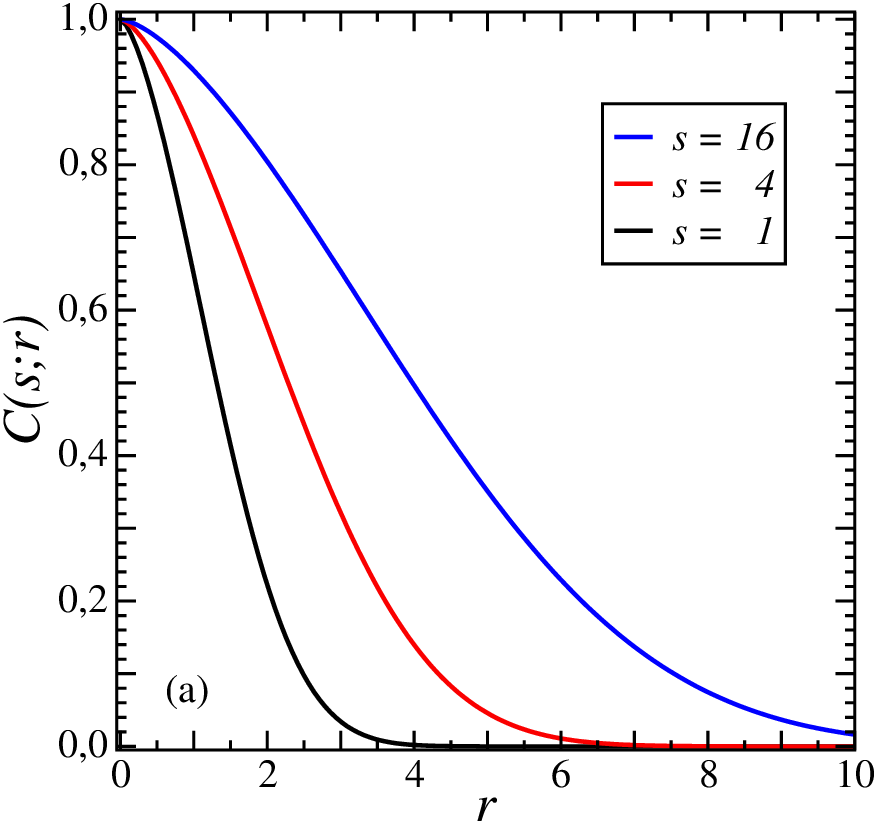}  ~
\includegraphics[width=0.3\hsize]{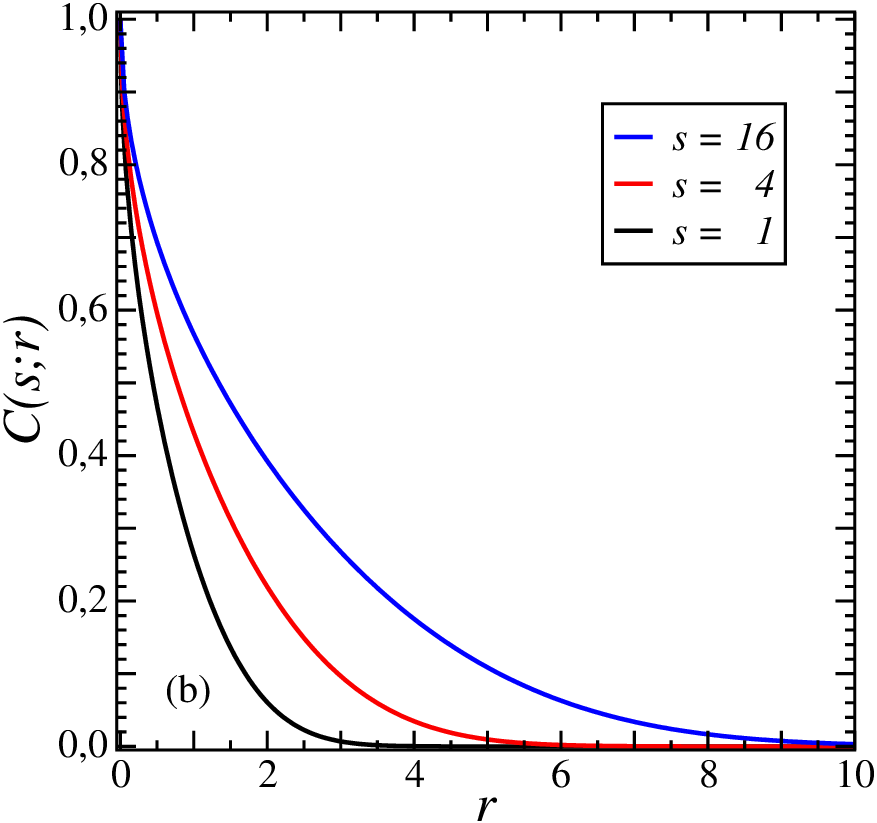}  ~
\includegraphics[width=0.3\hsize]{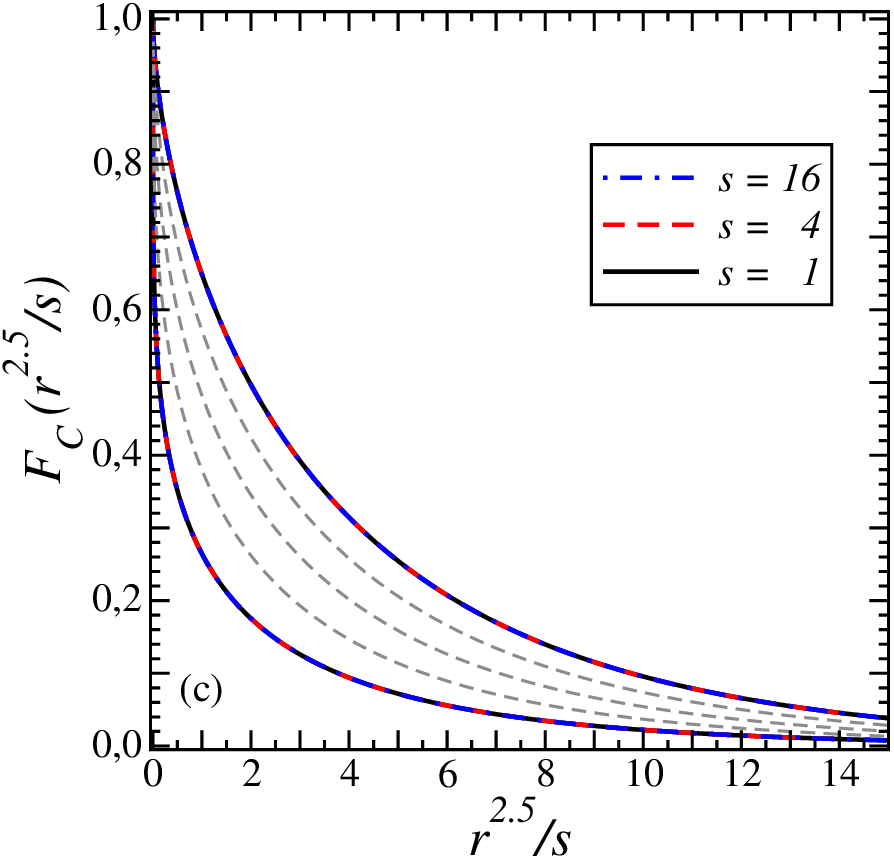}  ~
\caption[fig2]{Physical ageing in the single-time correlator (\ref{gl:voter-C1-bas}) of the voter model for $d<2+\theta$ dimensions, 
in the special case of a spectral index $\theta=\demi$.
Panel (a) shows $C(s;r)$ for $d=1$ and the three times $s=[1,4,16]$.
Panel (b) shows $C(s;r)$ for $d=2$ and the same three times $s$.
The data collapse of the associated scaling function $C(s;r)=F_C\bigl(1;r^{2.5}/s)$, when replotted as a function of $r^{2.5}/s$, 
is shown in  panel (c) where the upper curves correspond to $d=1$ and the lower curves to $d=2$. 
The gray dashed lines indicate the scaling function for $d=[1.25,1.50,1.75]$ from top to bottom. 
\label{fig2} }
\end{figure}
Figure~\ref{fig2} illustrates the three defining properties of ageing of the voter model on a fractal substrate with a spectral index $\theta=\demi$. 
Figure~\ref{fig2}a displays $C(s;r)$ as a function of $r$ for three values of $s$ and a geometric dimension $d=1$ 
of the substrate.\footnote{This case corresponds to the $1D$ Glauber-Ising model quenched to $T=0$.} 
With increasing values of $s$, the dynamics slows down, as evidenced by the curves moving
upwards and to the right. Since for different values of $s$, 
there are distinct curves, time-translation-invariance does not hold. Finally, figure~\ref{fig2}c shows the data collapse with
arises when the curves are replotted as a function of the variable $\bigl( \frac{r}{\ell(s)}\bigr)^{2+\theta}$ (upper curves). 
The same observations can be made for a geometric dimension $d=2$, as shown in figure~\ref{fig2}b. 
Their data collapse is also shown in figure~\ref{fig2}c (lower curves). 
Since many fractal substrates will have a geometric fractal dimension $1<d<2$, the corresponding scaling curves should fall in between those shown in figure~\ref{fig2}c 
(illustrated by the dashed gray lines for dimensions $d=[1.25,1.50,1.75]$). 


\subsection{Density of active interfaces} 
As a corollary, we obtain the asymptotic long-time scaling behaviour of the density $n_{\rm r}(t)=\demi\bigl(1-C(t;{\bf 1})\bigr)$ of reactive interfaces, 
where $\bf 1$ refers to the distance between nearest-neighbour sites. 
Since that distance is still within the scaling regime, we can expand (\ref{gl:voter-C1-bas}) around a nearest-neighbour separation $r\to 0$ and find, for $d_s<2$  
\BEQ \label{gl:alpha}
n_{\rm r}(t) \sim -\left.\frac{\partial C(t;{r})}{\partial {r}}\right|_{{r}=|\vec{1}|} \sim t^{-\alpha} \;\; , \;\; \alpha = 1-\frac{d}{2+\theta} = 1 - \frac{d_s}{2}
\EEQ
This power-law decay is consistent with the known result (\ref{gl:densact}) on euclidean lattices with $d<2$. 
In table~\ref{tab:2} we compare the prediction (\ref{gl:alpha}) with previous results of numerical studies \cite{Such06,Bab08} where indeed $d_s<2$ 
(these studies may be subject to important corrections due to finite-size effects and due to
finite iteration levels in the construction of the fractal). Their numerical results first reproduce the expected power-law decay for the enveloppe and second provide estimates
for $\alpha$. The values of $\alpha_{\rm num}$ are far from the euclidean prediction $\alpha_{\rm eucl}=1-d_f/2$ \cite{Such06,Bab08} 
but are seen to be in good agreement with (\ref{gl:alpha}). 

\begin{table}[tb]
\begin{center}
\begin{tabular}{|l|lll|c|}  \hline
name                       & ~$d_s$       & ~$\alpha_{\rm th}$ & ~$\alpha_{\rm num}$  & Ref. \\ \hline
Sierpinksi triangle        & $1.3652$     & $0.317$            & $0.3456(34)$         & \cite{Such06} \\[0.1cm] 
Sierpinksi carpet SC(3,3)  & $1.2818$     & $0.3591$           & $0.366(4)$           & \cite{Bab08} \\ 
Sierpinksi carpet SC(4,6)  & $1.3238$     & $0.3381$           & $0.345(1)$           & \cite{Bab08} \\ 
Sierpinksi carpet SC(5,10) & $1.3546$     & $0.3327$           & $0.329(1)$           & \cite{Bab08} \\ \hline
\end{tabular}\end{center}
\caption[tab2]{Test of the scaling relation $\alpha_{\rm th}=1-d_s/2$ on some fractal lattices. The spectral dimension $d_s$ is from table~\ref{tab:1}. 
}
\label{tab:2}
\end{table}

This successful comparison with long-standing simulational data may be viewed as an {\it a posteriori} 
confirmation of our treatment of fractal lattices, as proposed in section~\ref{sec:2}.3. 

\subsection{The case $d>2+\theta$}
For $d>2+\theta$ (or $d_s>2$), we have from (\ref{gl:b-exp}) that $b=(d-2-\theta)/(2+\theta)$. It is convenient to simplify the scaling function
$f_C(\mathfrak{u}) = \mathfrak{u}^{\phi} F\bigl( \mathfrak{u}^{2+\theta}\bigr)$. With the choice $\phi = 2+\theta-d<0$ and $\mathfrak{v}=\mathfrak{u}^{2+\theta}$ we find
\BEQ
\bigl(2+\theta\bigr)^2 \mathfrak{v} F''(\mathfrak{v}) + \left[ \bigl(2+\theta\bigr)\bigl(4+2\theta-d\bigr) + \mathfrak{v}\right] F'(\mathfrak{v}) =0
\EEQ
This is treated via two straightforward integrations as in the case before and  leads to 
\BEQ
F(\mathfrak{v}) = F_0 + F_1\, \Gamma\left( \frac{d}{2+\theta}-1, \frac{\mathfrak{v}}{(2+\theta)^2} \right) \;\; , \;\; d>2+\theta
\EEQ
Once more, the first boundary conditions leads to $F_0=0$. For small arguments $\mathfrak{v}\ll 1$, it must be recognised that the physical constraint $C(t;0)\stackrel{!}{=}1$ is outside
the scaling regime. Up to an arbitrary choice of a scaling amplitude $\mathfrak{C}_0$ we end up with 
\addtocounter{equation}{-4}   
\begin{subequations}          
\addtocounter{equation}{1}    
\BEQ \label{gl:voter-C1-haut} 
C(t;r) 
= \mathfrak{C}_0\, r^{2+\theta-d} \frac{\Gamma\bigl(\frac{d}{2+\theta}-1,\frac{1}{(2+\theta)^2}\frac{r^{2+\theta}}{t}\bigr)}{\Gamma\bigl(\frac{d}{2+\theta}-1\bigr)} \;\; ; \;\; d>2+\theta
\EEQ
\end{subequations}           
\addtocounter{equation}{3}   
\noindent
The two equations (\ref{gl:voter-C1}) together give the single-time correlator for all dimensions $d\ne 2+\theta$ and
will serve as input for the calculations of the other observables. 
On a non-fractal substrate with $\theta=0$ they reduce to the well-known exact result \cite{Henk26}.

However, the saturation seen in the density $n_{\rm r}(t)$ of active interfaces above the upper critical dimension 
is a non-scaling result and not within reach of the scaling solution discussed here. 

Also, we shall not deal explicitly with the case $d=2+\theta$ (or $d_s=2$). 
By analogy with the voter model on an euclidean $2D$ lattice, we expect a modified scaling ansatz with additional
logarithmic factors \cite{Frac97,Corb24b,Henk26} and the scaling function should be continuous in $d$. 

\subsection{Characteristic length}
An immediate consequence of (\ref{gl:voter-C1}) is the well-known scaling of the characteristic length scale $\ell(t)$. 
One of the several possibilities is to deduce it from the averages of the
correlation function 
\begin{align}
\ell^2(t) &= \bigl\langle r^2 \bigr\rangle(t) \:=\: 
\frac{\int_{\mathbb{R}^d} \!\D\vec{r}\; \vec{r}^2\, C(t;\vec{r})}{\int_{\mathbb{R}^d} \!\D\vec{r}\;  C(t;\vec{r})}
\:=\: \frac{\int_0^{\infty} \!\D r\: r^{d+1}\, 
      \Gamma\bigl(1-\frac{d}{2+\theta},\frac{1}{(2+\theta)^2} \frac{r^{2+\theta}}{t}\bigr)}{\int_0^{\infty} \!\D r\: r^{d-1}\, 
      \Gamma\bigl(1-\frac{d}{2+\theta},\frac{1}{(2+\theta)^2} \frac{r^{2+\theta}}{t}\bigr)}
\nonumber \\
&=    \frac{t^{(d+1)/(2+\theta)}}{t^{(d-1)/(2+\theta)}}\frac{\int_0^{\infty} \!\D \mathfrak{u}\: \mathfrak{u}^{d+1}\, 
      \Gamma\bigl(1-\frac{d}{2+\theta},\mathfrak{u}^{2+\theta}\bigr)}{\int_0^{\infty} \!\D \mathfrak{u}\: \mathfrak{u}^{d-1}\, 
      \Gamma\bigl(1-\frac{d}{2+\theta},\mathfrak{u}^{2+\theta}\bigr)}
\:\sim\: t^{2/(2+\theta)}
\label{gl:z-dyn}
\end{align}
here for $d<2+\theta$. An analogous end result is easily established for $d>2+\theta$ as well. 
It follows that $\theta$ can be identified as the exponent which describes the anomalous sub-diffusion in this model, for all spatial dimensions $d$. 
Clearly, we recover from (\ref{gl:z-dyn}) the expected dynamical exponent $\mathpzc{z}=2+\theta$. 
This is very much in line with long-standing expectations \cite{Shau85a,Shau85b} described in section~\ref{sec:2}, based on the diffusion equation. 

\section{Two-time auto-correlator} \label{sec:4}

Applying again the rates (\ref{gl:2.2}), the enveloppe of the two-time correlator $C(t,s;r)$ is found by solving the differential equation
\BEQ \label{gl:2tC}
\partial_t C(t,s;\vec{r}) = \demi \Delta_{\vec{r}} C(t,s;\vec{r}) \;\; , \;\; C(s,s;\vec{r}) = C(s;\vec{r})
\EEQ
where the equal-time case serves as an initial condition. In the continuum limit, spatial rotation-invariance implies that $\vec{r}\mapsto r = |\vec{r}|$. 
On a fractal with spectral index $\theta$, one has explicitly
\BEQ \label{gl:4.2} 
\partial_t C(t,s;r) = \demi \frac{1}{r^{d-1}} \frac{\partial}{\partial r}\left( r^{d-1-\theta} \frac{\partial C(t,s;r)}{\partial r}\right) \;\; , \;\;
C(s,s;r) = C(s;r)
\EEQ

We shall again look for a scaling description where simultaneously $t,s\to \infty$ and $r\to\infty$ such that the scaling variables
\BEQ \label{gl:4.3}
y = \frac{t}{s} \;\; , \;\; u = \frac{r^{2+\theta}}{s}
\EEQ
are kept fixed. Later on, we shall focus on the two-time auto-correlator $C(ys,s)=C(ys,s;0)$. 

\subsection{Case $d<2+\theta$}
We begin with the case $d<2+\theta$, or equivalently $d_s<2$ where $b=0$. 
Then we can write the scaling ansatz for the two-time correlator $C(t,s;r) = F(y;u)$ and the scaling function must be found from
\BEA \label{gl:Fscal}
\partial_y F(y;u) &=& \frac{(2+\theta)^2}{2} \left( \frac{d}{2+\theta} \frac{\partial}{\partial u} + u \frac{\partial^2}{\partial u^2} \right) F(y;u) = 0 \;\; ; \;\; \\
F(1;u)            &=& f_C\bigl(\mathfrak{u}^{2+\theta}\bigr) \:=\: \frac{\Gamma\bigl(1-\frac{d_s}{2},\frac{u}{(2+\theta)^2} \bigr)}{\Gamma\bigl(1-\frac{d_s}{2}\bigr)}
\EEA
where we recall the scaling function $f_C$ for the single-time correlator from section~\ref{sec:3}, 
keeping in mind the definitions (\ref{gl:2.5},\ref{gl:4.3}) of the scaling variables $\mathfrak{u}$ and $u$, respectively. 
This scaling ansatz works since eq.~(\ref{gl:Fscal}), and by consequence its solution $F(y;u)$ as well, has become independent of the waiting time $s$. 

One of the standard methods to solve linear partial differential equations is the method of separation of variables, e.g. \cite{Boas06}. 
With the ansatz $F(y;u) = Y(y) U(u)$ we obtain the separate equations
\BEQ \label{gl:sepa}
Y'(y) = -\frac{(2+\theta)^2}{2} K^2 Y(y) \;\; , \;\;
u U''(u) + \frac{d}{2+\theta} U'(u) + K^2 U(u) = 0
\EEQ
which will be used to generate a sufficiently large basis of solutions that the sought solution can be found by linear combination of these by summing over the admissible
values of the separation constant $K$. The first of these functions is trivially an exponential. 
The second one is found via a Frob\'enius series: this proceeds via the ansatz $U(u) = \sum_{n=0}^{\infty} a_n u^{n+\upsilon}$ with $a_0\ne 0$. Inserting 
this into (\ref{gl:sepa}) gives a recursion for the $a_n$ and fixes $\upsilon$. We merely quote the result
\begin{subequations}
\begin{align}
Y(y) &= y_1 \exp\left[ - \frac{(2+\theta)^2}{2} K^2 \bigl(y-1\bigr)\right] \\
U(u) &= u_0 \Gamma\left(\frac{d}{2+\theta}\right) J_{-1+d/(2+\theta)}\left( 2 K \sqrt{u\,}\right) \left( K \sqrt{u\,}\right)^{1-d/(2+\theta)} \nonumber \\
&~~+ u_1 \Gamma\left(2-\frac{d}{2+\theta}\right) J_{1-d/(2+\theta)}\left( 2 K \sqrt{u\,}\right) \left( \frac{K}{\sqrt{u\,}}\right)^{-1+d/(2+\theta)} 
\end{align}
\end{subequations}
with the Bessel functions $J_{\pm p}(x)$ and the constants $u_{0,1}$ and $y_1$. These parts can be combined to give a formal presentation of the solution
\BEA
F(y;u) &=& \int_0^{\infty} \!\D K\: e^{-\frac{(2+\theta)^2}{2} K^2 (y-1)} 
\left[ F_0(K^2) \left( K \sqrt{u\,}\,\right)^{1-d/(2+\theta)} J_{-1+d/(2+\theta)}\left( 2K \sqrt{u\,}\,\right) \right. \nonumber \\
& & \left. ~~
     + F_1(K^2) \left( \frac{\sqrt{u\,}}{K}\right)^{1-d/(2+\theta)} J_{1-d/(2+\theta)}\left( 2K \sqrt{u\,}\,\right) \right]
\nonumber \\
&=&  \int_0^{\infty} \!\D K\: 2K\, e^{-\frac{(2+\theta)^2}{2} K^2 (y-1)} 
\left[ \mathscr{F}_0(K^2) \left( K \sqrt{u\,}\,\right)^{1-d/(2+\theta)} J_{-1+d/(2+\theta)}\left( 2K \sqrt{u\,}\,\right) \right. \nonumber \\
& & \left. ~~ +\mathscr{F}_1(K^2) \left( K \sqrt{u\,}\,\right)^{1-d/(2+\theta)} J_{1-d/(2+\theta)}\left( 2K \sqrt{u\,}\,\right) \right]
\EEA
(below, it will turn out to be better to use $\mathscr{F}_{0,1}(K^2)$ rather than the more immediate $F_{0,1}(K^2)$). 

In this work, we are mainly interested in the two-time auto-correlator $C(ys,s)$, 
which with the low-argument expansion \cite[(9.1.10)]{Abra65} of the Bessel functions can be obtained as
\BEA
\lefteqn{ C(ys,s) = F(y;0) =  \lim_{u\to 0} \int_0^{\infty} \!\D \kappa\: e^{-\frac{(2+\theta)^2}{2} \kappa (y-1)} \times} \nonumber \\
& & \times
\left[ \mathscr{F}_0(\kappa) \frac{\bigl(\sqrt{\kappa\, u\,}\,\bigr)^{\frac{d}{2+\theta}-1}}{\Gamma(\frac{d}{2+\theta})} \bigl(\sqrt{\kappa\,u\,}\,\bigr)^{1-\frac{d}{2+\theta}}
+ \mathscr{F}_1(\kappa) \frac{\bigl(\sqrt{\kappa\,u\,}\,\bigr)^{1-\frac{d}{2+\theta}}}{\Gamma(2-\frac{d}{2+\theta})} \bigl({\sqrt{\kappa\,u\,}}\,\bigr)^{1-\frac{d}{2+\theta}} 
+\ldots \right]
\nonumber \\
&=& \lim_{u\to 0} \int_0^{\infty} \!\D \kappa\: e^{-\frac{(2+\theta)^2}{2} \kappa (y-1)}  \left[ \frac{1}{\Gamma(\frac{d}{2+\theta})} \mathscr{F}_0(\kappa) + {\rm o}(u) \right]
\nonumber \\
&=& \frac{1}{\Gamma(\frac{d}{2+\theta})} \bigl( \lap{\mathscr{F}_0(\kappa)}\bigr)\left(\frac{(2+\theta)^2}{2}(y-1)\right)
\label{gl:4.9}
\EEA
with the Laplace transform $\lap{f}(s) =\mathscr{L}\bigl(f(t)\bigr)(s) := \int_0^{\infty} \!\D t\: e^{-st} f(t)$. 
Hence the two-time auto-correlator is essentially the Laplace transform of the initial condition $\mathscr{F}_0$ 
but does not contain explicitly the function $\mathscr{F}_1$. 
The long-time asymptotics of $C(ys,s)$, for $y\gg 1$,  might be found via a Tauberian theorem \cite{Fell71} 
for which the behaviour of $\mathscr{F}_0(\kappa)$ for small arguments $\kappa\ll 1$ would be required.

The two yet undetermined functions $\mathscr{F}_{0,1}(K^2)$ 
must be found from the initial condition, valid for all $u\in\mathbb{R}_+$ where from now one we shall use the abbreviation $p=1-\frac{d}{2+\theta}$
\BEA
\lefteqn{F(1;u) = \frac{\Gamma(p, \frac{1}{(2+\theta)^2} u)}{\Gamma(p)} }  \label{eq:4.8} \\
&=& \int_0^{\infty} \!\!\D K\: 2K 
\left[ \mathscr{F}_0(K^2) \left( K \sqrt{u\,}\,\right)^{p} J_{-p}\left( 2K \sqrt{u\,}\,\right) 
     + \mathscr{F}_1(K^2) \left( K {\sqrt{u\,}}\,\right)^{p} J_{p}\left( 2K \sqrt{u\,}\,\right) \right] \nonumber
\EEA
In what follows, we shall change the integration variable to $\kappa=K^2$.

The functions $\mathscr{F}_{0,1}(\kappa)$ should be found from the orthogonality relations of the Bessel functions $J_{\pm p}(x)$. 
For reference, we reproduce in appendix~A the standard way of calculating their orthogonality relations on a finite interval, following \cite{Boas06}. 
But we shall rather need their orthogonality relations for a semi-infinite interval, which for $0<|p|<1$ read
\begin{subequations}
\begin{align}
\int_0^{\infty} \!\D u\; u\, J_p(uR) J_p(u r)    &= \frac{1}{R} \delta(R-r)  \\
\int_0^{\infty} \!\D u\: u\, J_p(uR) J_{-p}(ur)  &= \cos(\pi p) \frac{1}{R} \delta(R-r) + \frac{2\sin \pi p}{\pi} \frac{\bigl(\frac{R}{r}\bigr)^{p}}{R^2 - r^2} 
\end{align}
\end{subequations}
where $\delta$ is the Dirac distribution \cite{Schw1950,Gelf64}.  
The first of these is well-known from the litt\'erature \cite{Watson1922}, the second one is proven in appendix~B. 
We shall require them here in the form
\begin{subequations} \label{gl:Bessel-ortho}
\begin{align}
\int_0^{\infty} \!\D u\; J_p(2K \sqrt{u\,}\,) J_p(2M \sqrt{u\,}\,)     &= \frac{1}{2K} \delta(K-M)  \label{gl:Bessel-ortho-pp} \\
\int_0^{\infty} \!\D u\: J_p(2K \sqrt{u\,}\,) J_{-p}(2M \sqrt{u\,}\,)  &= \frac{\cos(\pi p)}{2K} \delta(K-M) + \frac{\sin \pi p}{\pi} \frac{\bigl(\frac{K}{M}\bigr)^{p}}{K^2 - M^2} 
\label{gl:Bessel-ortho-pm}
\end{align}
\end{subequations}
In order to apply these to the constraint (\ref{eq:4.8}) following from the initial conditions, we first define
\BEQ
g_{\pm}(M^2) := \int_0^{\infty} \!\D u\: F(1;u)\, u^{-p/2} J_{\pm p}\bigl(2M \sqrt{u\,}\,\bigr) \;\; , \;\; p := 1-\frac{d}{2+\theta} \in (0,1)
\EEQ
and then obtain via (\ref{gl:Bessel-ortho}) the two separate conditions
\begin{subequations} \label{gl:4.13}
\begin{align}
M^{-p} g_{-}(M^2) &= \mathscr{F}_0(M^2) +\cos \pi p\, \mathscr{F}_1(M^2) + \frac{\sin \pi p}{\pi} \int_0^{\infty} \!\D K\: 2K \left(\frac{K}{M}\right)^{2p} \frac{\mathscr{F}_1(K^2)}{K^2-M^2} \\
M^{-p} g_{+}(M^2) &= \cos\pi p\, \mathscr{F}_0(M^2) +\mathscr{F}_1(M^2)  + \frac{\sin \pi p}{\pi} \int_0^{\infty} \!\D K\: 2K \frac{\mathscr{F}_0(K^2)}{M^2-K^2} 
\end{align}
\end{subequations}
which by the new integration variables $\kappa=K^2$, $\mu=M^2$ can be cast into the form
\begin{subequations} \label{gl:4.14}
\begin{align}
\mu^{-p/2} g_{-}(\mu) &= \mathscr{F}_0(\mu) +\cos \pi p \, \mathscr{F}_1(\mu) 
                                + \frac{\sin \pi p}{\pi} \int_0^{\infty} \!\D \kappa\: \left(\frac{\kappa}{\mu}\right)^{p} \frac{\mathscr{F}_1(\kappa)}{\kappa-\mu} \\
\mu^{-p/2} g_{+}(\mu) &= \cos\pi p \, \mathscr{F}_0(\mu) +\mathscr{F}_1(\mu) 
                                + \frac{\sin \pi p}{\pi} \int_0^{\infty} \!\D \kappa\:  \frac{\mathscr{F}_0(\kappa)}{\mu-\kappa} 
\end{align}
\end{subequations}
These can be decoupled via a Mellin transformation (see appendix~C). For brevity of notation, we define
\begin{subequations} \label{gl:4.15}
\begin{align}
\mathscr{G}_{-}(u) &:= \mathscr{M}\left( \mu^{-p/2} g_{-}(\mu) \right)(u) \:=\: 
~~\frac{\bigl(2+\theta\bigr)^{2(1-u)}}{\Gamma(p)} \frac{\Gamma(u-p) \Gamma\bigl(1+p-u\bigr)}{\Gamma(2-u)}
\label{gl:4.15a} \\
\mathscr{G}_{+}(u) &:= \mathscr{M}\left( \mu^{-p/2} g_{+}(\mu) \right)(u) \:=\: 
-\frac{\bigl(2+\theta\bigr)^{2(1-u)}}{\Gamma(p)} \frac{\Gamma(u-1) \Gamma\bigl(1+p-u\bigr)}{\Gamma(1+p-u)}
\label{gl:4.15b} 
\end{align}
\end{subequations}
where the explicit calculation is carried out in appendix~D. Mellin-transforming, we then find, using also eqs.~(\ref{gl.C4},\ref{gl.C5}), 
that (\ref{gl:4.14}) turns into the following inhomogeneous linear system 
\BEQ \label{gl:4.16}
\left( \begin{array}{cc} 1                                & \cos\pi p + \sin \pi p \cot\pi(u-p) \\
                         \cos\pi p - \sin \pi p \cot\pi u & 1 \end{array} \right)
\left( \begin{array}{c} \mathscr{M}\bigl(\mathscr{F}_0\bigr)(u) \\ \mathscr{M}\bigl(\mathscr{F}_1\bigr)(u) \end{array} \right)
= \left( \begin{array}{c} \mathscr{G}_-(u) \\ \mathscr{G}_+(u) \end{array} \right)
\EEQ
Because of the identity $\bigl(\cos\pi p + \sin \pi p \cot\pi(u-p)\bigr)\bigl(\cos\pi p - \sin \pi p \cot\pi u\bigr)=1$, 
the determinant of the $2\times 2$ matrix in (\ref{gl:4.16}) vanishes. 
In order to assess the number of solutions of the singular system (\ref{gl:4.16}) 
one may find the kernel ${\cal K}$ by solving the corresponding homogeneous equation, e.g. \cite{Nobl88}. 
The kernel ${\cal K}$ can be cast as an arbitrary multiple of the vector 
\BEQ
\vec{k} = \left( \begin{array}{c} -1 \\ \cos\pi p +\sin \pi p \cot \pi (u-p) \end{array} \right)
\EEQ
The right-hand vector in (\ref{gl:4.16}) is in its orthogonal complement ${\cal K}^{\perp}$, since
\BEA
\lefteqn{\vec{k} \cdot \left( \begin{array}{c} \mathscr{G}_-(u) \\ \mathscr{G}_+(u) \end{array} \right) 
= \left( \begin{array}{c} -1 \\ \cos\pi p +\sin \pi p \cot \pi (u-p) \end{array} \right) \cdot \left( \begin{array}{c} \mathscr{G}_-(u) \\ \mathscr{G}_+(u) \end{array} \right) }
\nonumber \\
&=& - \mathscr{G}_-(u)+\bigl( \cos\pi p + \sin \pi p \cot\pi(u-p)\bigr)\mathscr{G}_+(u) \:=\: 0
\EEA
because of the identity (\ref{gl:D.4}). This means that (\ref{gl:4.16}) has infinitely many non-zero solutions \cite{Nobl88}. 
Since the auto-correlator (\ref{gl:4.9}) only depends on $\mathscr{F}_0(\kappa)$, we make the (arbitrary~!) choice 
\BEQ \label{gl:F1_0}
\mathscr{F}_1(\kappa)=0
\EEQ
If this is admitted, as we shall do from now on, it implies from (\ref{gl:4.16}) 
$\mathscr{M}\bigl(\mathscr{F}_0\bigr)(u)=\mathscr{G}_-(u)$, or because of the identity property of the Mellin transformation \cite{Flaj95}
\BEQ
\mathscr{F}_0(\kappa) = \kappa^{-p/2} g_-(\kappa) 
\EEQ
\begin{figure}[tb]
\includegraphics[width=0.3\hsize]{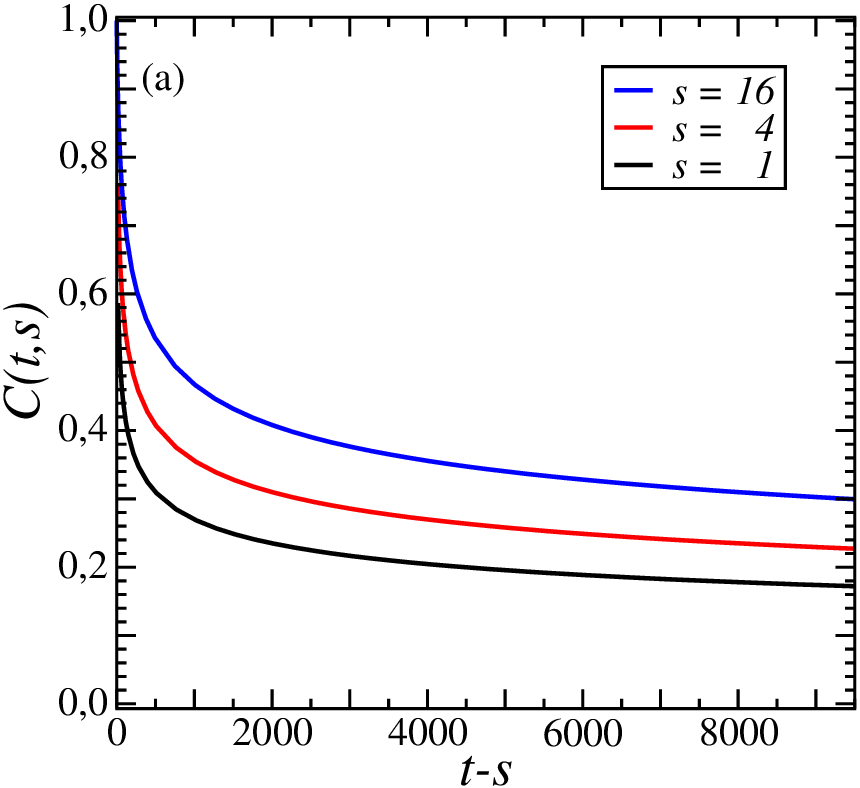}  ~
\includegraphics[width=0.3\hsize]{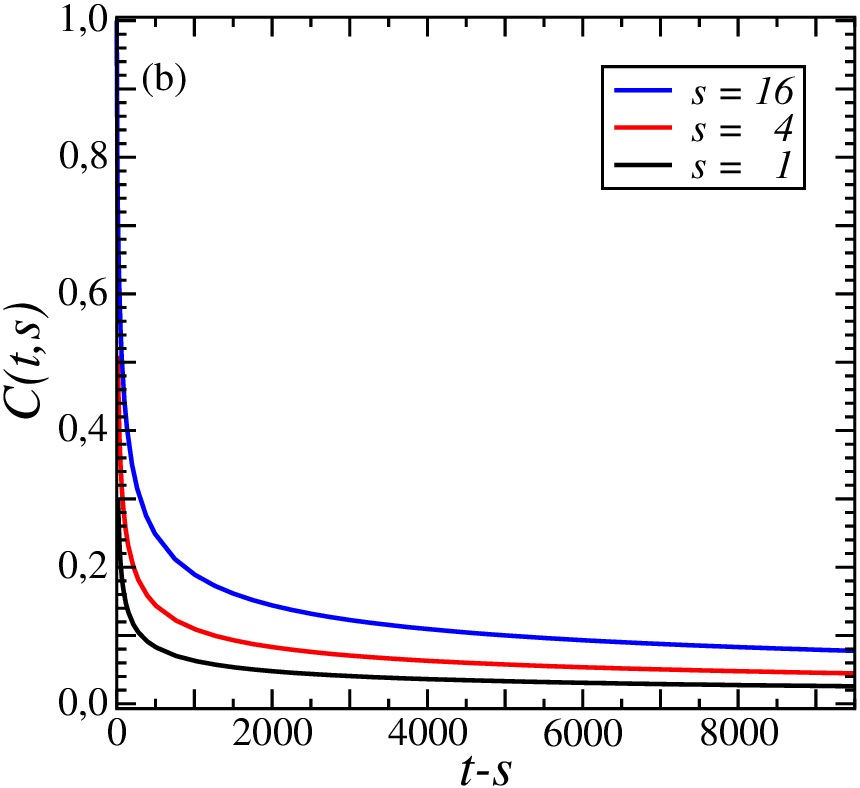}  ~
\includegraphics[width=0.3\hsize]{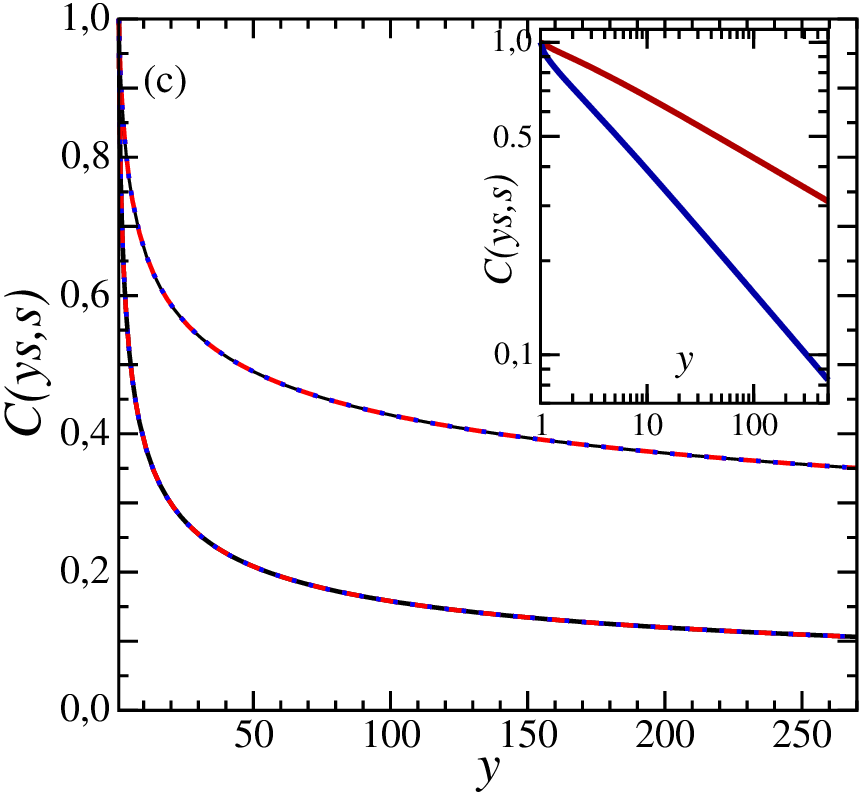}  ~
\caption[fig3]{Physical ageing in the two-time correlator (\ref{gl:4.23}) of the voter model for $d<2+\theta$ dimensions, 
in the special case of a spectral index $\theta=\demi$.
Panel (a) shows $C(t,s)$ over against $\tau=t-s$ for $d=1$ and the three times $s=[1,4,16]$.
Panel (b) shows similarly $C(t,s)$ for $d=2$ and the same three times $s$.
The data collapse of the associated scaling function $C(ys,s)=f_C(y)$, over against $y$, 
is shown in  panel (c). The upper curves correspond to $d=1$ and the lower curves to $d=2$. 
The almost straight lines in the log-log plot of the inset illustrate the algebraic decay for $y\gg 1$. 
\label{fig3} }
\end{figure}
For the calculation of the two-time auto-correlator $C(ys,s)$ via (\ref{gl:4.9}), we rather need its Laplace-transform (see again the appendix~D for the calculation)
\BEA
\left(\lap{\mathscr{F}_0(\kappa)}\right)(\sigma) 
&=& \frac{1}{\Gamma(p)} \left( \frac{\sigma}{(2+\theta)^{2}}\right)^{p-1} \frac{\Gamma(1-p)}{\Gamma(2-p)}\, {}_2F_1\left( 1-p,1;2-p;-\frac{(2+\theta)^2}{\sigma} \right)
\label{gl:4.21}
\EEA
which we give here 
as a hypergeometric function \cite{Abra65}. From this, and (\ref{gl:4.9}), the auto-correlator becomes
\BEQ \label{gl:4.23}
C(ys,s) =   \frac{1}{\Gamma\bigl(1-\frac{d_s}{2}\bigr)\Gamma\bigl(1+\frac{d_s}{2}\bigr)} \left( \frac{y-1}{2}\right)^{-d_s/2}  
{}_2F_1\left( \frac{d_s}{2},1;1+\frac{d_s}{2};-\frac{2}{y-1} \right)
\EEQ
where we find it convenient to express the scaling function in terms of the spectral dimension $d_s$. For an euclidean lattice, $d_s$ reduces to $d$ and we then recover the
exactly known result \cite[(2.19)]{Henk26}. Standard identities also imply that $C(s,s)=1$, 
as it should be and this provides a test for the correctness of our auxiliary hypothesis (\ref{gl:F1_0}). 

Examples for the two-time correlator are displayed in figure~\ref{fig3}. For spatial dimensions $d=1$ and $d=2$, we observe that the three defining properties of ageing are obeyed:
(i) slow dynamics, since with increasing waiting times $s$, the dynamics slows down and the curves in figure~\ref{fig3}ab move upwards and to the right, 
(ii) breaking of time-translation-invariance since there are distinct curves for different values of $s$ and
(iii) dynamical scaling as evidenced by the data collapse in figure~\ref{fig3}c.  

The asymptotic behaviour $C(ys,s)\sim y^{-\lambda/\mathpzc{z}}$ for $y\gg 1$ (see the inset in figure~\ref{fig3}c) 
permits to extract the autocorrelation exponent, and we read off $\frac{\lambda}{\mathpzc{z}}=\frac{d_s}{2}$. 
Since the single-time correlator has already produced in section~\ref{sec:3} the result $\mathpzc{z}=2+\theta$, the exact autocorrelation exponent is
\BEQ
\lambda = \frac{d_s}{2} (2+\theta) = \frac{d_s}{2} \frac{2 d}{d_s} = d
\EEQ
which equals in the voter model the geometric fractal dimension $d$, as it would be the case on an euclidean lattice. 
This surprisingly simple finding is the main result of this section, together with the explicit scaling function (\ref{gl:4.23}). 
The fractal nature of the substrate becomes
apparent in the value $\mathpzc{z}=2+\theta$ of the dynamical exponent, controlled by the value of the spectral index. 

In spite of the rather simple result, the above calculation has it also made clear that the voter model is non-trivial as put into evidence by the non-trivial form of its scaling functions. 

\subsection{Case $d>2+\theta$}
In the case $d>2+\theta$ (or equivalently $d_s>2$), we have $b=\bigl(d-2-\theta\bigr)/\bigl(2+\theta\bigr)$. This leads to the slightly modified scaling form
\BEQ \label{gl:4.25}
C(ys,s;r) = s^{-b} F(y;u) = r^{-(d-2-\theta)} E(y;u) \;\; ; \;\; y = \frac{t}{s} \;\; , \;\; u = \frac{r^{2+\theta}}{s} 
\EEQ
and we also define a new scaling function $E(y;u)$, generalising (\ref{gl:voter-C1-haut}). The best we can do here is to find the finite amplitude $E(y;0)$. 

Since the waiting time $s$ does not enter into the differential equation (\ref{gl:4.2}), we can repeat the above calculation and find again, 
with the only change that $p=\frac{d}{2+\theta}-1>0$ takes a new meaning
\BEA 
\partial_y F(y;u) &=& \frac{(2+\theta)^2}{2} \left( \frac{d}{2+\theta} \frac{\partial}{\partial u} + u \frac{\partial^2}{\partial u^2} \right) F(y;u) = 0 \label{gl:Escal}  \\
F(1;u)            &=& u^{-p} \frac{\Gamma\bigl(p,\frac{u}{(2+\theta)^2} \bigr)}{\Gamma\bigl(p\bigr)} \:=\: u^{-p} E(1;u) \label{gl:Einit} 
\EEA
The technique of separation of variables can be applied as before and the result is conveniently formulated with the scaling function $E(y;u)=u^p F(y;u)$ 
\BEA
E(y;u) &=&  \int_0^{\infty} \!\D K\: 2K\, e^{-\frac{(2+\theta)^2}{2} K^2 (y-1)} 
\left[ \mathscr{F}_0(K^2) \left( K \sqrt{u\,}\,\right)^{p} J_{p}\left( 2K \sqrt{u\,}\,\right) \right. \nonumber \\
& & \left. ~~ +\mathscr{F}_1(K^2) \left( K \sqrt{u\,}\,\right)^{p} J_{-p}\left( 2K \sqrt{u\,}\,\right) \right]
\EEA
with the initial condition, taken from (\ref{gl:Einit}) 
\BEA
E(1;u) &=& \int_0^{\infty} \!\!\D K\: 2K 
\left[ \mathscr{F}_0(K^2) \left( K \sqrt{u\,}\,\right)^{p} J_{p}\left( 2K \sqrt{u\,}\,\right) 
     + \mathscr{F}_1(K^2) \left( K {\sqrt{u\,}}\,\right)^{p} J_{-p}\left( 2K \sqrt{u\,}\,\right) \right] \nonumber \\ && \label{gl:4.29}
\EEA
It can also be shown that now the small-argument limit becomes
\BEQ \label{gl:4.30}
E(y;0) = \lim_{u\to 0} u^p F(y;u) = \frac{1}{\Gamma(1-p)} \bigl(\lap{ \mathscr{F}_1(\kappa)}\bigr)\left( \frac{(2+\theta)^2}{2} \bigl(y-1\bigr)\right)
\EEQ
We remark that with respect to the previous sub-section, the r\^oles of $\mathscr{F}_0$ and $\mathscr{F}_1$ have been interchanged. 
The initial condition (\ref{gl:4.29}) is recast into two separate equations via the Bessel orthogonality relations (\ref{gl:Bessel-ortho}).  
We can define the functions $g_{\pm}(\kappa)$ as in (\ref{gl:4.13}), but with $F(1;u)$ replaced by $E(1;u)$. We then obtain
\begin{subequations} \label{gl:4.31}
\begin{align}
\mu^{-p/2} g_{-}(\mu) &= \mathscr{F}_1(\mu) +\cos \pi p \, \mathscr{F}_0(\mu) 
                                + \frac{\sin \pi p}{\pi} \int_0^{\infty} \!\D \kappa\: \left(\frac{\kappa}{\mu}\right)^{p} \frac{\mathscr{F}_0(\kappa)}{\kappa-\mu} \\
\mu^{-p/2} g_{+}(\mu) &= \cos\pi p \, \mathscr{F}_1(\mu) +\mathscr{F}_0(\mu) 
                                + \frac{\sin \pi p}{\pi} \int_0^{\infty} \!\D \kappa\:  \frac{\mathscr{F}_1(\kappa)}{\mu-\kappa} 
\end{align}
\end{subequations}
As before, a Mellin transformation reduces this to a singular $2\times 2$ system with an infinite number of solutions. Because of (\ref{gl:4.30}), the choice
\BEQ
\mathscr{F}_0(\kappa)=0
\EEQ
appears natural and implies
\BEQ
\mathscr{F}_1(\kappa) = \kappa^{-p/2} g_{-}(\kappa)   
\EEQ
We recall from appendix~D its Laplace transform 
\BEQ
\bigl(\lap{ \mathscr{F}_1(\kappa)}\bigr)\bigl(\sigma\bigr) 
= \frac{\Gamma(1-p)}{\Gamma(p)\Gamma(2-p)} \left( \frac{\sigma}{(2+\theta)^{2}}\right)^{p-1}  {}_2F_1\left( 1-p,1;2-p;-\frac{(2+\theta)^2}{\sigma} \right)
\EEQ
but we remind the reader that now $p=\frac{d_s}{2}-1=\frac{d}{2+\theta}-1$. 
This produces the amplitude 
\BEA
E(y;0) &=& \lim_{r\to 0} r^{d-2-\theta}\, C(ys,s;r) \nonumber \\
&=& \frac{1}{\Gamma(\frac{d_s}{2}-1)\Gamma(3-\frac{d_s}{2})}\left(\frac{2}{y-1}\right)^{d_s/2-2}\, {}_2F_1\left(2-\frac{d_s}{2},1;3-\frac{d_s}{2};-\frac{2}{y-1}\right) 
\label{gl:4.35}
\EEA
which is positive for $y>1$ and related to the correlator via (\ref{gl:4.25}). Again, we find that this scaling expression only depends on the spectral dimension $d_s$.

The non-triviality of the voter model resides in its correlation functions, to be discussed further in section~\ref{sec:5}.

\section{Discussion} \label{sec:5}

We have studied the non-equilibrium dynamics of the voter model on a fractal substrate.  
In turn, this is characterised by its geometric fractal dimension $d=d_f$, its spectral index $\theta$ or alternatively its spectral dimension $d_s$ 
(the knowledge of two of these parameters allows to find the third, see (\ref{gl:d_spectral}), and we implicitly restricted to cases where the consideration of
two fractal dimensions $d_f, d_s$ is sufficient (rather than up to six, e.g. \cite{Pati23})). 
We used a phenomenological approach, based on the close relationship of the voter model with the diffusion equation,.
The topological properties of the fractal substrate were introduced via a space-dependent diffusion constant ${\cal D}(r)\sim r^{-\theta}$. 
Thereby, we restrict attention to the {\em enveloppes} of the correlation functions which are smooth functions of their temporal and spatial coordinates 
\cite{Shau85a,Shau85b} and exclude from consideration any log-periodic oscillation around these. 
Then, in principle, the behaviour of the voter model on any specific fractal can be obtained by selecting the appropriate values of $d$ and $\theta$. 
The admissibility of arbitrary values $d,\theta\in\mathbb{R}_+$ relies on the applicability of a scaling ansatz to characterise the time-space behaviour 
of single-time and two-time correlators and which had been used before to obtain these correlation functions for $d\in\mathbb{R}_+$ and $\theta=0$ \cite{Henk26}. 

\begin{enumerate} 
\item In general, our result for the enveloppes of both single-time and two-time correlators satisfy the requirements of physical ageing, for any values of $d,\theta$, see
figures~\ref{fig2},~\ref{fig3}.  
The scaling variables are found to depend on the spectral index $\theta$. 
However, the scaling functions (\ref{gl:voter-C1},\ref{gl:4.23},\ref{gl:4.35}) turn out to depend on the spectral dimension $d_s$ only. 
This further implies that the euclidean results are
included as special cases for $\theta=0$. Is this a peculiarity of the voter model or rather a generic feature~? 
\item In particular, for $d<2+\theta$ (or $d_s<2$) we find the exponents
\BEQ \label{gl:5.1}
\mathpzc{z} = 2 +\theta \;\; , \;\; \lambda = d
\EEQ
The values of $\lambda$ interpolate smoothly between the euclidean lattices when $d\in\mathbb{N}$, which reminds one of analogous observations for the
critical exponents $\Theta,\delta$ in the critical contact process, see figure~\ref{fig1}ab. 
On the other hand, no such smooth interpolation seems to exist for $\mathpzc{z}$, as also suggested in figure~\ref{fig1}c for the critical contact process. 

The reason for this different behaviour remains unknown. 
\item For geometric dimensions $d<2+\theta$ (or $d_s<2$), 
the single-time correlator implies a long-time power-law decay for the enveloppe of the density $n_{\rm r}(t)$ of active interfaces
\BEQ \label{gl:5.2}
n_{\rm r}(t) \sim t^{-\alpha} \;\; ; \;\; \alpha = 1 - \frac{d_s}{2}
\EEQ
This prediction can be compared with the long-standing results of numerical simulations \cite{Such06,Bab08} and precise calculations of $d_s$ for several deterministic fractals
\cite{Ramm83,Dasg99,Schu00,Fran26} and a nice agreement is found, see table~\ref{tab:2}. First, this is an empirical {\it a posteriori} 
confirmation of the phenomenological approach used here,
second it resolves an open issue about interpreting the value of $\alpha$ \cite{Such06,Bab08} and
third, it shows that the origin of the decay of $n_{\rm r}(t)$ is topological. Further tests of (\ref{gl:5.2}) would be welcome. 

For $d>2+\theta$, the small-distance regime where (\ref{gl:5.2}) holds is no longer in the time-space scaling region studied here. 
\item On the technical side, notably obtaining the two-time auto-correlator did require to find new orthogonality relations (\ref{gl:Bessel-ortho}) between the Bessel functions 
$J_{\pm p}(x)$ on a semi-infinite interval (which still await a mathematically rigorous proof). 
Via a subsequent Mellin transformation, the consistency with the single-time correlator led to a singular system of equations,
with an infinite number of solutions, from which by physical intuition we selected the most plausibly-looking one which is continuous in the euclidean limit $d_s\to d$. 
Can one improve on this technique of solution and avoid the apparent arbitrariness~?
\item On fractals, the motion of particles is always sub-diffusive, see (\ref{gl:5.1}), which looks intuitive because of the `holes' with respect to the euclidean lattices.  
What would happen if a long-range voter model could be analysed on fractal substrates, since long-range interactions may in certain cases lead to super-diffusive motion 
\cite{Corb24a,Corb24b,Corb24e}~? 
\item Motivated by the empirical finding that the aggressivity of tumors increases with the fractal dimension $d_f$ \cite{Cros97,Elki22}, 
recent studies on tumor growth took the fractal nature of tissues into account, via a space-dependent diffusion constant, 
and emphasised the importance of the spectral dimension $d_s$ on growth patterns \cite{Fume25,Faja26}. 
While the voter model itself certainly has no relevance in tumour growth, the present work is also intended as a conceptual background for this ongoing work 
and the idea of ageing scaling forms might become of predictive value in relevant models. 
\item What could be said about the {\em quantum} dynamics of the voter model, 
notably since it may be viewed as a system of Ising spins coupled to two distinct baths at zero and infinite temperature, respectively \cite{Droz89}~? 
\end{enumerate}

\noindent
{\bf Acknowledgements:}
Remerciements chaleureuses \`a S. Fumeron, A. L\'opez, T. Sandev, M. Olmo Fajardo et Z. Khaled pour des discussions fructueuses.  
Herzlichen Dank an A. Franz f\"ur hilfreiche Korrespondenz. 
This work was supported by the french ANR-PRME UNIOPEN (ANR-22-CE30-0004-01).

\newpage 
\appsection{A}{On Bessel orthogonal relations I} 

\addtocounter{equation}{-1}
The mutual orthogonality relations of the Bessel functions $J_{\pm p}(x)$ and $Y_p(x)$ are studied on the finite interval $[0,1]$. 
A well-known identity concerns the zeros $\alpha,\beta$ of the Bessel function $J_p(\alpha)=J_p(\beta)=0$ and reads \cite{Boas06}
\BEQ \label{gl.A0}
\int_0^1 \!\D x\; x\, J_p(\alpha x) J_p(\beta x) 
= \left\{ \begin{array}{ll} \demi J_{p+1}^2(\alpha) & \mbox{\rm\small ~~;~ if $\alpha=\beta$} \\ 0 & \mbox{\rm\small ~~;~ if $\alpha\ne \beta$} \end{array} \right. 
\EEQ
We are interested in the mutual orthogonality of the other Bessel functions with the squared order $p^2$. Our results are given in the form of three lemmas. 

\noindent
{\bf Lemma A.1:} {\it If 
$p$ is not an integer and if $\alpha,\beta\in\mathbb{R}$ are the non-vanishing zeros of the Bessel functions $J_{\pm p}(x)$ such that
$J_p(\alpha) = J_{-p}(\beta)=0$, then one  has the identity}
\BEQ \label{gl.A1}
\int_0^1 \!\D x\;  x\, J_p(\alpha x) J_{-p}(\beta x) 
                                                     = \frac{2\sin p\pi}{\pi} \frac{1}{\alpha^2 -\beta^2} \left( \frac{\alpha}{\beta}\right)^p 
\EEQ

\noindent
{\bf Proof:} Since $p\not\in\mathbb{Z}$, the zeros $\alpha\neq\beta$ are always distinct and are taken to be non-vanishing. The integral (\ref{gl.A1}) always exists at its lower limit. 
We follow \cite{Boas06}, and let $u(x):=J_p(\alpha x)$ and $v(x):=J_{-p}(\beta x)$. 
One has from Bessel's differential equation $x\bigl(x u'\bigr)'+\bigl(\alpha^2 x^2 - p^2\bigr)u=0$ and similarly for $v$ with $\alpha$ replaced by $\beta$. 
Multiplying the first equation by $v$, the second one with $u$, subtracting these two equations and then integrating over the interval $[0,1]$, one obtains \cite{Boas06} 
\BD
\biggl. \left[ v(x) x u'(x) -u(x) x v'(x) \right] \biggr|_0^1  +\bigl(\alpha^2 - \beta^2\bigr) \int_0^1 \!\D x\; x\, u(x) v(x) = 0 \tag{*} 
\ED
One must now evaluate the two boundary terms $T_{1,2}$ in (*). The first one simply is
\BD
T_1 = v(1)\, 1 u'(1) - u(1)\, 1 v'(1) = J_{-p}(\beta) {J_p}'(\alpha) - J_p(\alpha) {J_{-p}}'(\beta) = 0
\ED
because of the definition  of $\alpha,\beta$ as zeros of $J_{\pm p}(x)$, respectively. 
For the second term, we recall the small-argument behaviour of the $J_{\pm p}(x)$ \cite{Abra65} and consider 
\BEA 
\lefteqn{T_2 = \lim_{x\to 0} \left[ J_{-p}(\beta x) x {J_p}'(\alpha x) - J_p(\alpha x) x {J_{-p}}'(\beta x) \right]} 
\nonumber \\
&=& \lim_{x\to 0} \left[ \left( \frac{\beta x}{2}\right)^{-p} \!\frac{x}{\Gamma(1-p)} \left( \left(\frac{\alpha x}{2}\right)^p \frac{1}{\Gamma(1+p)} \right)' 
- \left( \frac{\alpha x}{2}\right)^{p} \frac{x}{\Gamma(1+p)} \left( \left(\frac{\beta x}{2}\right)^{-p} \frac{1}{\Gamma(1-p)} \right)' \,\right]
\nonumber \\
&=& \lim_{x\to 0} \left[ \frac{1}{\Gamma(1-p)\Gamma(1+p)} \left( \frac{\alpha}{\beta}\right)^p \left( x^{-p+1+p-1} p - x^{p+1-p-1} \bigl(-p\bigr) \right) \right] 
\:=\: \frac{2p}{\Gamma(1-p)\Gamma(1+p)} \left(\frac{\alpha}{\beta}\right)^p
\nonumber
\EEA
Inserting this into the basic identity (*) and recalling that $\Gamma(1-p)\Gamma(1+p)=p \pi /\sin p\pi$ \cite[(6.1.17)]{Abra65} gives the assertion. \hfill q.e.d. 

\noindent 
{\bf Remark:} one might as well define $\alpha,\beta$ as zeros of the derivatives of the Bessel functions, viz. ${J_p}'(\alpha)={J_{-p}}'(\beta)=0$ 
and repeat the argument essentially unchanged. Although the two Bessel functions $J_{p}(x)$ and $J_{-p}(x)$ are distinct for $p\not\in\mathbb{Z}$, they are not mutually orthogonal. 

\newpage
\noindent
{\bf Lemma A.2:} {\it If $0<p<1$, and $\alpha,\beta\in\mathbb{R}$ are the non-vanishing and distinct zeros of the Bessel functions $Y_p(x)$ such that
$Y_p(\alpha) = Y_{p}(\beta)=0$ with $\alpha\neq\beta$, then one  has the identity}
\BEQ
\int_0^1 \!\D x\;  x\, Y_p(\alpha x) Y_{p}(\beta x) = -\frac{2}{\pi}\cot\bigl(\pi p\bigr) \frac{\bigl(\alpha/\beta)^p - \bigl(\beta/\alpha\bigr)^p}{\alpha^2 - \beta^2} 
\EEQ

\noindent
{\bf Remark:} in contrast to $J_p(x)$, see eq.~(\ref{gl.A0}), this is non-vanishing for $\alpha\ne\beta$. 

\noindent
{\bf Proof:} the zeros $\alpha,\beta$ of $Y_p(x)$ are always taken to be distinct and non-vanishing. Now let $u(x):=Y_p(\alpha x)$ and $v(x):=Y_{p}(\beta x)$. 
As before in the proof of Lemma A.1, one obtains the basic relation (*) derived before in the proof of Lemma A.1 \cite{Boas06}. 
The evaluation of the first boundary term is immediate 
\BD
T_1 = v(1)\, 1 u'(1) - u(1)\, 1 v'(1) = Y_{p}(\beta) {Y_p}'(\alpha) - Y_p(\alpha) {Y_{p}}'(\beta) = 0
\ED
because of the definition  of $\alpha,\beta$. For the second one, we use the definition of $Y_p(x)$ in terms of the $J_{\pm p}(x)$ \cite[(9.1.2)]{Abra65}
\BD
Y_p(x) = \frac{1}{\sin \pi p} \left( J_p(x) \cos\bigl( \pi p\bigr) - J_{-p}(x) \right)
\ED
and can then write by taking into account the small-argument-behaviour of the $J_{\pm p}(x)$ \cite{Abra65}, where the second-order contribution must be included for the $J_{-p}(x)$
\BEA
\lefteqn{T_2 = \lim_{x\to 0} \biggl[ x \biggl( Y_p(\beta x) {Y_p}'(\alpha x) - Y_p(\alpha x) {Y_p}'(\beta x) \biggr) \biggr]}
\nonumber \\
&=& \frac{1}{\sin^2 \pi p} \lim_{x\to 0} x \biggl[ \left(J_p(\beta x)\cos(\pi p)-J_{-p}(\beta x)\right)\left({J_p}'(\alpha x)\cos(\pi p)-{J_{-p}}'(\alpha x)\right) \biggr. 
\nonumber \\
&& \biggl. ~~~~- \left(J_p(\alpha x)\cos(\pi p)-J_{-p}(\alpha x)\right)\left({J_p}'(\beta x)\cos(\pi p)-{J_{-p}}'(\beta x)\right)  \biggr]
\nonumber \\
&=& \frac{1}{\sin^2 \pi p} \lim_{x\to 0} x  \left\{ 
\left[ \frac{\bigl(\beta x/2\bigr)^p}{\Gamma(1+p)}\left( \frac{\bigl(\alpha x/2\bigr)^p}{\Gamma(1+p)} \right)' 
     - \frac{\bigl(\alpha x/2\bigr)^p}{\Gamma(1+p)}\left( \frac{\bigl(\beta x/2\bigr)^p}{\Gamma(1+p)} \right)'\, \right] \cos^2 \pi p   \right. 
\nonumber \\
& & + \left. \left[ \frac{\bigl(\alpha x/2\bigr)^p}{\Gamma(1+p)}\left( \frac{\bigl(\beta x/2\bigr)^{-p}}{\Gamma(1-p)} \left( 1 - \frac{\bigl(\beta x/2\bigr)^2}{1-p}\right)\right)'
    + \frac{\bigl(\alpha x/2\bigr)^{-p}}{\Gamma(1-p)}\left( 1 - \frac{\bigl(\alpha x/2\bigr)^2}{1-p}\right)\left( \frac{\bigl(\beta x/2\bigr)^{p}}{\Gamma(1+p)} \right)'
\right.  \right. 
\nonumber \\
& & \left. \left. ~~- \frac{\bigl(\beta x/2\bigr)^p}{\Gamma(1+p)}\left( \frac{\bigl(\alpha x/2\bigr)^{-p}}{\Gamma(1-p)} \left( 1 - \frac{\bigl(\alpha x/2\bigr)^2}{1-p}\right)\right)'
    - \frac{\bigl(\beta x/2\bigr)^{-p}}{\Gamma(1-p)}\left( 1 - \frac{\bigl(\beta x/2\bigr)^2}{1-p}\right)\left( \frac{\bigl(\alpha x/2\bigr)^{p}}{\Gamma(1+p)} \right)' \,
\right] \cos \pi p \right. 
\nonumber \\
& & \left. + \left[ \frac{\bigl(\beta x/2\bigr)^{-p}}{\Gamma(1-p)}\left( 1 - \frac{\bigl(\beta x/2\bigr)^2}{1-p}\right) 
      \left( \frac{\bigl(\alpha x/2\bigr)^{-p}}{\Gamma(1-p)} \left( 1 - \frac{\bigl(\alpha x/2\bigr)^2}{1-p}\right)\right)' \right. \right.
\nonumber \\
& & \left. \left. 
    ~~- \frac{\bigl(\alpha x/2\bigr)^{-p}}{\Gamma(1-p)} \left( 1 - \frac{\bigl(\alpha x/2\bigr)^2}{1-p}\right) 
      \left( \frac{\bigl(\beta x/2\bigr)^{-p}}{\Gamma(1-p)}\left( 1 - \frac{\bigl(\beta x/2\bigr)^2}{1-p}\right) \right)' \, \right]  \right\} 
\nonumber 
\EEA
Evaluating this to the leading non-trivial order for $x\ll 1$ gives further
\BEA
T_2 &=& \frac{1}{\sin^2 \pi p} \lim_{x\to 0} \left\{ \frac{\cos^2 \pi p}{\Gamma^2(1+p)} \left( \frac{\alpha\beta}{4}\right)^p 
\left[ p x^{1+p+p-1} - p x^{1+p+p-1}\right] \left( 1+ {\rm O}(x) \right) \right. 
\nonumber \\
& & \left. + \frac{\cos \pi p}{\Gamma(1+p)\Gamma(1-p)} 
\left\{ \left(\frac{\alpha}{\beta}\right)^p 
\left[ x^{1+p} \left( -p x^{-p-1} - \frac{\beta^2}{4}\frac{2-p}{1-p}x^{1-p}\right) - x^{1-p}\left( 1 - \frac{\beta^2/4}{1-p} x^2\right) p x^{p-1} \right] \right. \right.
\nonumber \\
& & \left. \left. ~~+\left(\frac{\beta}{\alpha}\right)^p \left[ x^{1-p} \left( 1 - \frac{\alpha^2/4}{1-p} x^2\right) p x^{p-1} 
-x^{1+p} \left( -p x^{-p-1} - \frac{\alpha^2}{4}\frac{2-p}{1-p}  x^{1-p}\right) \right] \right\}  \right.
\nonumber \\
& & + \left. \frac{1}{\Gamma^2(1-p)}\left(\frac{4}{\alpha\beta}\right)^p 
\left[ x^{1-p}\left( 1 - \frac{\beta^2/4}{1-p} x^2\right)\left( -p x^{-p-1}-\frac{\alpha^2}{4}\frac{2-p}{1-p}x^{1-p}\right)
\right. \right. 
\nonumber \\
& & \left. \left. ~~-x^{1-p}\left( 1 - \frac{\alpha^2/4}{1-p} x^2\right)\left( -p x^{-p-1}-\frac{\beta^2}{4}\frac{2-p}{1-p} x^{1-p}\right)\right] \right\}
\nonumber \\
&=& \frac{1}{\sin^2 \pi p} \lim_{x\to 0} \left\{ {\rm O}(x^{2p}, x^{2p+1}) \right.
\nonumber \\
& & \left. 
+\frac{\cos \pi p}{\Gamma(1-p)\Gamma(1+p)}\left\{ 
\left(\frac{\alpha}{\beta}\right)^p\left[ -2p x^0 + {\rm O}(x^2)\right] + \left(\frac{\beta}{\alpha}\right)^p\left[ -2p x^0 + {\rm O}(x^2)\right] \right\} 
\right.
\nonumber \\
& & \left. 
+\frac{1}{\Gamma^2(1-p)} \left( \frac{4}{\alpha\beta}\right)^p \left[ -p x^{-2p} +\frac{\beta^2}{4} \frac{p}{1-p} x^{2-2p} + p x^{-2p} - \frac{\alpha^2}{4} \frac{p}{1-p} x^{2-2p} 
+{\rm O}(x^{3-2p}) \right] \right\}
\nonumber \\
&=& \frac{1}{\sin^2 \pi p} \left\{ \frac{\cos\bigl(\pi p\bigr) 2p}{\Gamma(1-p)\Gamma(1+p)}\left[ \left(\frac{\beta}{\alpha}\right)^p - \left(\frac{\alpha}{\beta}\right)^p \right]
\left( 1 + {\rm O}(x^2)\right) \right. 
\nonumber \\
& & \left. + \frac{1}{\Gamma^2(1-p)}\left(\frac{4}{\alpha\beta}\right)^p \demi \frac{\beta^2-\alpha^2}{1-p} x^{2-2p} \left( 1 + {\rm O}(x)\right) \right\}
\nonumber
\EEA
Since $0\leq p<1$, the second term in the last line, of order $x^{2-2p}$n will disappear for $x\ll 1$. Therefore, if $\alpha\ne \beta$ are non-vanishing and $0\leq p<1$, we have
\BD
\lim_{x\to 0} x\bigl( Y_p(\beta x){Y_p}'(\alpha x)-Y_p(\alpha x){Y_p}'(\beta x) \bigr) = \frac{2p \cos \pi p}{\sin^2 \pi p}\frac{1}{\Gamma(1-p)\Gamma(1+p)} 
\left[ \left(\frac{\beta}{\alpha}\right)^p - \left(\frac{\alpha}{\beta}\right)^p \right]
\ED
Finally, inserting this into (*), with the identity  \cite[(6.1.17)]{Abra65}, we arrive at the assertion. \hfill q.e.d.

\newpage
\noindent
{\bf Lemma A.3:} {\it If $0<p<1$ and if $\alpha,\beta\in\mathbb{R}$ are the non-vanishing zeros of the Bessel functions $J_{p}(x)$ and $Y_p(x)$ such that
$J_p(\alpha) = Y_{p}(\beta)=0$, then one  has the identity}
\BEQ
\int_0^1 \!\D x\;  x\, J_p(\alpha x) Y_{p}(\beta x) = -\frac{2}{\pi}\left(\frac{\alpha}{\beta}\right)^p \frac{1}{\alpha^2 - \beta^2} 
\EEQ

\noindent
{\bf Proof:} since $J_p(x)$ and $Y_p(x)$ are independent functions, $\alpha$ and $\beta$ are always distinct. 
Letting $u(x) = J_p(\alpha x)$ and $v(x) =Y_p(\beta x)$, one derives as
before the relation (*) \cite{Boas06}. The first boundary term is
\BD
T_1 = v(1)\, 1 u'(1) - u(1)\, 1 v'(1) = Y_{p}(\beta) {J_p}'(\alpha) - J_p(\alpha) {Y_{p}}'(\beta) = 0
\ED
and the second boundary term $T_2$ becomes with the definition \cite[(9.1.2)]{Abra65} of $Y_p(x)$ 
\BEA
\lefteqn{T_2 = \lim_{x\to 0} \biggl[ x \biggl( Y_p(\beta x) {J_p}'(\alpha x) - J_p(\alpha x) {Y_p}'(\beta x) \biggr) \biggr]}
\nonumber \\
&=& \frac{1}{\sin\pi p} \lim_{x\to 0} x \biggl[ 
\bigl(J_p(\beta x)\cos(\pi p)-J_{-p}(\beta x)\bigr){J_p}'(\alpha x)- J_p(\alpha x)\bigl({J_p}'(\beta x)\cos(\pi p)-{J_{-p}}'(\beta x)\bigr)  
\biggr]
\nonumber \\
&=& \frac{1}{\sin\pi p} \lim_{x\to 0} x \left\{ 
\left[ \frac{\bigl(\beta x/2\bigr)^p}{\Gamma(1+p)} \left(\frac{\bigl(\alpha x/2\bigr)^p}{\Gamma(1+p)}\right)' 
- \frac{\bigl(\beta x/2\bigr)^p}{\Gamma(1+p)} \left(\frac{\bigl(\alpha x/2\bigr)^p}{\Gamma(1+p)}\right)'\, \right] \cos\pi p  \right. 
\nonumber \\
& & \left. ~~+\left[ \frac{\bigl(\alpha x/2\bigr)^p}{\Gamma(1+p)} \left( \frac{\bigl(\beta x/2\bigr)^{-p}}{\Gamma(1-p)}\left( 1 -\frac{\bigl(\beta x/2\bigr)^2}{1-p}\right) \right)' 
-  \frac{\bigl(\beta x/2\bigr)^{-p}}{\Gamma(1-p)}\left( 1 -\frac{\bigl(\beta x/2\bigr)^2}{1-p}\right) \left( \frac{\bigl(\alpha x/2\bigr)^p}{\Gamma(1+p)} \right)' \, \right] \right\}
\nonumber \\
&=& \frac{1}{\sin\pi p} \lim_{x\to 0} \left\{ {\rm O}(x^{2p+1}) - \left(\frac{\alpha}{\beta}\right)^p \frac{2p}{\Gamma(1-p)\Gamma(1+p)} x^0 + {\rm O}(x) \right\} 
\nonumber \\
&=& - \frac{2}{\pi} \left( \frac{\alpha}{\beta}\right)^p
\nonumber
\EEA
where in the last line the Gamma function identity \cite[(6.1.17)]{Abra65} was used again. Insertion into (*) gives the assertion. \hfill q.e.d. 

\noindent
{\bf Remark:} All new identities in this appendix have also been checked numerically. 

\newpage
\appsection{B}{On Bessel orthogonal relations II} 

The mutual orthogonality relations of the Bessel functions $J_{\pm p}(x)$ and $Y_p(x)$ are studied on the semi-infinite interval $[0,\infty)$. It is not possible to glean
these results by simply rescaling those for a finite interval derived in appendix~A, although a comparison may be of heuristic value. 

We begin by recalling an identity due to Watson \cite[p. 465, eq. (1)]{Watson1922}, in a slightly recast form 
\begin{subequations} 
\BEQ \label{gl.B1.a}
\int_0^{\infty} \!\!\D R\: R\, F(R)\! \int_0^{\infty} \!\!\D u\: u  \mathscr{C}_p(u R) \mathscr{C}_p(ur) = F(r) - \frac{2}{\pi} \frac{\sin \alpha \sin(\alpha+p\pi)}{\sin \pi p}
\int_0^{\infty} \!\!\D R\: R\, F(R)\, \frac{\bigl(\frac{R}{r}\bigr)^{p} - \bigl(\frac{r}{R}\bigr)^p}{R^2 - r^2}
\EEQ
where $F(R)$ is a conveniently chosen test function such that all integrals in question exist, $0\leq p \leq \demi$ and
\BEQ
\mathscr{C}_p(z) = \cos \alpha\, J_p(x) + \sin \alpha\, Y_p(x)
\EEQ
\end{subequations}
Notice that the last factor in (\ref{gl.B1.a}) is symmetric under the exchange $R\leftrightarrow r$, as it should be. 

We shall now consider three special choices for $\alpha$ and notice the identities which follow. 
First, we let $\alpha=0$ and note the resulting identity with the help of
a Dirac distribution $\delta$ \cite{Schw1950,Gelf64}, namely
\begin{subequations} \label{gl.B2}
\BEQ \label{gl.B2.a}
\int_0^{\infty} \!\D u\; u\, J_p(uR) J_p(u r) = \frac{1}{R} \delta(R-r) = \frac{1}{r}\delta(R-r)
\EEQ
and rigorously proven at least for $p\geq -\demi$ \cite[p. 456, eq. (1)]{Watson1922}.
This shortens the notation since we need no longer write a test function $F(R)$ explicitly. 
Second, we let $\alpha=\frac{\pi}{2}$. With $\sin\bigl(\frac{\pi}{2}+p \pi\bigr)=\cos p \pi$, one has
\BEQ \label{gl.B2.b}
\int_0^{\infty} \!\D u\; u\, Y_p(uR) Y_p(u r) = \frac{1}{R} \delta(R-r) - \frac{2}{\pi} \cot p\pi \frac{\bigl(\frac{R}{r}\bigr)^{p} - \bigl(\frac{r}{R}\bigr)^p}{R^2 - r^2}
\EEQ
Third, we let $\alpha=\frac{\pi}{4}$, use $\sin\bigl(\frac{\pi}{4}+p \pi\bigr)= \sin\frac{\pi}{4}\cos p \pi + \cos\frac{\pi}{4} \sin p \pi$ and read off
\BEA
\lefteqn{ \demi \int_0^{\infty} \!\D u\: u\, \bigl( J_p(uR) + Y_p(uR) \bigr) \bigl( J_p(ur) + Y_p(ur) \bigr)} \nonumber \\
&=& \frac{1}{R} \delta(R-r) - \frac{1+\cot p\pi}{\pi} \frac{\bigl(\frac{R}{r}\bigr)^{p} - \bigl(\frac{r}{R}\bigr)^p}{R^2 - r^2} \nonumber
\EEA
Developing the product in the left-hand side of this equation and then using the relations (\ref{gl.B2.a}) and (\ref{gl.B2.b}) leads to
\BEQ \label{gl.B2.c}
\demi \int_0^{\infty} \!\D u\: u\, \bigl( J_p(uR) Y_p(u r) + J_p(u r) Y_p(u R) \bigr) = - \frac{1}{\pi} \frac{\bigl(\frac{R}{r}\bigr)^{p} - \bigl(\frac{r}{R}\bigr)^p}{R^2 - r^2}
\EEQ
\end{subequations}
To this, we add a forth relation, as follows. Recall the definition $Y_p(x)= \bigl( J_p(x)\cos(p\pi) - J_{-p}(x)\bigr)/\sin p \pi$, valid for $p$ non-integer \cite{Abra65}, 
which is inserted into (\ref{gl.B2.c}) to yield
{\small\BD
\frac{1}{2\sin\pi p}\int_0^{\infty} \!\!\D u\: u \bigl[ J_p(uR) \bigl( J_p(ur)\cos(p \pi) - J_{-p}(ur)\bigl) + J_p(ur) \bigl( J_p(uR)\cos(p \pi) - J_{-p}(uR)\bigl) \bigr] 
= -\frac{1}{\pi} \frac{\bigl(\frac{R}{r}\bigr)^{p} - \bigl(\frac{r}{R}\bigr)^p}{R^2 - r^2}
\ED}
and developing the product turns this into
\BD
\frac{1}{2\sin\pi p}\int_0^{\infty} \!\D u\: u \bigl[ J_p(uR) J_p(ur) 2\cos(p \pi) - J_p(uR) J_{-p}(ur) - J_p(ur) J_{-p}(uR) \bigr] 
= -\frac{1}{\pi} \frac{\bigl(\frac{R}{r}\bigr)^{p} - \bigl(\frac{r}{R}\bigr)^p}{R^2 - r^2} 
\ED
which is further simplified via (\ref{gl.B2.a}) to the final form
\addtocounter{equation}{-1}
\begin{subequations}
\addtocounter{equation}{3}
\BEQ \label{gl.B2.d}
\demi \int_0^{\infty} \!\!\D u\: u \bigl[ J_p(uR) J_{-p}(ur) + J_p(ur)J_{-p}(uR) \bigr] = \cos(\pi p) \frac{1}{R} \delta(R-r) 
+ \frac{\sin \pi p}{\pi} \frac{\bigl(\frac{R}{r}\bigr)^{p} - \bigl(\frac{r}{R}\bigr)^p}{R^2 - r^2}
\EEQ
where in the last equation we must explicitly admit that $p$ is non-integer. 
\end{subequations}

All these rigorous relations (\ref{gl.B2}) are symmetric under the exchange $R\leftrightarrow r$. In order to proceed beyond that restriction, we formulate

\noindent 
{\bf Lemma B.1} {\it For $p$ non-integer, one has for the Bessel functions $J_{\pm p}(x)$ the identities}
\begin{subequations} \label{gl.B3}
\begin{align}
\demi \int_0^{\infty} \!\!\D u\: u\, \bigl[ J_p(uR) J_{-p}(ur) + J_p(ur)J_{-p}(uR) \bigr] &= \frac{\cos(\pi p)}{R}  \delta(R-r) 
+ \frac{\sin \pi p}{\pi} \frac{\bigl(\frac{R}{r}\bigr)^{p} - \bigl(\frac{r}{R}\bigr)^p}{R^2 - r^2} \label{gl.B3.a}
\\
\demi \int_0^{\infty} \!\!\D u\: u\, \bigl[ J_p(uR) J_{-p}(ur) - J_p(ur)J_{-p}(uR) \bigr] &=
\frac{\sin \pi p}{\pi} \frac{\bigl(\frac{R}{r}\bigr)^{p} + \bigl(\frac{r}{R}\bigr)^p}{R^2 - r^2}   \label{gl.B3.b}
\end{align}
\end{subequations}
As an immediate consequence, we add the two identities (\ref{gl.B3}) and obtain

\noindent 
{\bf Lemma B.2} {\it For $p$ non-integer, one has for the Bessel functions $J_{\pm p}(x)$ the identity}
\BEQ \label{gl.B4}
\int_0^{\infty} \!\D u\: u\,  J_p(uR) J_{-p}(ur)  = \cos(\pi p) \frac{1}{R} \delta(R-r) 
+ \frac{2\sin \pi p}{\pi} \frac{\bigl(\frac{R}{r}\bigr)^{p}}{R^2 - r^2} 
\EEQ
Eq.~(\ref{gl.B4}) is the sought orthogonality relation between $J_p(x)$ and $J_{-p}(x)$, required in the main text. For an integer $p=n\in\mathbb{Z}$, this reduces to (\ref{gl.B2.a}).  

Clearly, the identities (\ref{gl.B3},\ref{gl.B4}) are to be interpreted as distributions \cite{Schw1950,Gelf64}. 
Since the existing rigorous results (\ref{gl.B2}) only cover the symmetric case, we shall first
present a heuristic technique in order to re-derive eq.~(\ref{gl.B3.a}) and shall then use this same technique to obtain (\ref{gl.B3.b}) as well. 

\noindent
{\bf Heuristic proof:} Our argument starts from the well-known identity \cite[(11.3.29)]{Abra65} for a non-integer $p$ 
\begin{align}
& \bigl( K^2 - L^2\bigr) \int^z \!\D t\: t\, \mathscr{C}_p(K t) \mathscr{D}_{-p}(Lt) \nonumber \\
&=  z\biggl( K \mathscr{C}_{p+1}(Kz) \mathscr{D}_{-p}(Lz) - L\mathscr{C}_p(Kz) \mathscr{D}_{-p+1}(Lz)\biggr) -2p\, \mathscr{C}_p(Kz) \mathscr{D}_{-p}(Lz)  \tag{*}
\nonumber
\end{align} 
where the integral is considered as a function of the upper limit $z$,  $\mathscr{C}_p(z), \mathscr{D}_{-p}(z)$ are cylinder functions of order $\pm p$, 
respectively, and $K,L$ are constants. \\

\noindent \underline{Step 1.} In order to recover (\ref{gl.B3.a}), we recast (*) as follows
\begin{align}
& {\demi \int_0^{\infty} \!\D u\: u \biggl( J_p(uR) J_{-p}(ur) + J_{-p}(uR) J_p(ur) \biggr) \bigl( R^2 - r^2\bigr) }
\nonumber \\
&= \lim_{z\to\infty} \left\{ \frac{z}{2} \biggl( R J_{p+1}(Rz) J_{-p}(rz) - r J_{p}(Rz) J_{-p+1}(rz) \biggr) -p J_{p}(Rz) J_{-p}(rz) \right. 
\nonumber \\
&  \left. ~~~~~~+  \frac{z}{2} \biggl( R J_{p}(rz) J_{-p+1}(Rz) -r J_{p+1}(rz) J_{-p}(Rz) \biggr) +p J_{-p}(Rz) J_{p}(rz) \right\} 
\nonumber \\
&~~~- \lim_{z\to 0} \left\{ \frac{z}{2} \biggl( R J_{p+1}(Rz) J_{-p}(rz) - r J_{p}(Rz) J_{-p+1}(rz) \biggr) -p J_p(Rz) J_{-p}(rz) \right. 
\nonumber \\
&  \left. ~~~~~~+  \frac{z}{2} \biggl( R J_{-p+1}(Rz) J_{p}(rz) - r J_{-p}(Rz) J_{p+1}(rz) \biggr) +p J_{-p}(Rz) J_{p}(rz) \right\} 
\nonumber \tag{\#}
\end{align} 
such that the sought limit decomposes into  ${\cal L}=T_1 - T_2$ which are found separately. 
The evaluation of the second term $T_2$ is analogous to previous estimates, as presented in appendix~A, and relies on the leading term
in the formal series expansion of the $J_{\pm p}(x)$
\BEA
\lefteqn{T_2 = \lim_{z\to 0} \left\{ \frac{z}{2}\left[ R\frac{\bigl(Rz/2\bigr)^{p+1}}{\Gamma(2+p)}\frac{\bigl(r z/2\bigr)^{-p}}{\Gamma(1-p)} 
-r \frac{\bigl(Rz/2\bigr)^{p}}{\Gamma(1+p)}\frac{\bigl(r z/2\bigr)^{-p+1}}{\Gamma(2-p)} \right] -p\frac{\bigl(Rz/2\bigr)^{p}}{\Gamma(1+p)}\frac{\bigl(rz/2\bigr)^{-p}}{\Gamma(1-p)}
\right. }\nonumber \\
& & \left. ~~~~~+\frac{z}{2} \left[  -r\frac{\bigl(rz/2\bigr)^{p+1}}{\Gamma(2+p)}\frac{\bigl(R z/2\bigr)^{-p}}{\Gamma(1-p)} 
+R \frac{\bigl(rz/2\bigr)^{p}}{\Gamma(1+p)}\frac{\bigl(R z/2\bigr)^{-p+1}}{\Gamma(2-p)} \right] +p\frac{\bigl(rz/2\bigr)^{p}}{\Gamma(1+p)}\frac{\bigl(Rz/2\bigr)^{-p}}{\Gamma(1-p)} \right\}
\nonumber \\
&=& \lim_{z\to 0} \demi\left\{ {\rm O}(z^2) -\frac{2p}{\Gamma(1-p)\Gamma(1+p)} \left(\frac{R}{r}\right)^p + {\rm O}(z^2) + \frac{2p}{\Gamma(1-p)\Gamma(1+p)} \left(\frac{r}{R}\right)^p \right\}
\nonumber \\
&=& -\frac{\sin \pi p}{\pi} \left[ \left(\frac{R}{r}\right)^p - \left(\frac{r}{R}\right)^p \right]
\nonumber 
\EEA
and we re-used \cite[(6.1.17)]{Abra65}. 

For the formal evaluation of $T_1$, we require the leading large-argument asymptotics 
$J_p(z)\simeq \sqrt{\frac{2}{\pi z}\,}\cos\bigl(z-\frac{p\pi}{2}-\frac{\pi}{4}\bigr)$ \cite{Abra65} and
also use the trigonometric identity \cite[(4.3.32)]{Abra65}. Then we can formally estimate the leading terms  
{\small\BEA
\lefteqn{\hspace{-0.9cm}T_1 = \lim_{z\to\infty} \left\{ \frac{z}{2} \left[ R \sqrt{\frac{2}{\pi Rz}\,}\cos\bigl(Rz-\frac{(p+1)\pi}{2}-\frac{\pi}{4}\bigr)
                                                              \sqrt{\frac{2}{\pi rz}\,}\cos\bigl(rz-\frac{(-p)\pi}{2}-\frac{\pi}{4}\bigr) \right. \right.}
\nonumber \\
& & \left. \left.     ~~~~~~~~ - r \sqrt{\frac{2}{\pi Rz}\,}\cos\bigl(Rz-\frac{p\pi}{2}-\frac{\pi}{4}\bigr)
                                   \sqrt{\frac{2}{\pi rz}\,}\cos\bigl(rz-\frac{(-p+1)\pi}{2}-\frac{\pi}{4}\bigr) \right] \right.
\nonumber \\
& & \left.            ~~~~~~    -2p\sqrt{\frac{2}{\pi Rz}\,}\cos\bigl(Rz-\frac{p\pi}{2}-\frac{\pi}{4}\bigr)\sqrt{\frac{2}{\pi rz}\,}\cos\big(rz-\frac{(-p)\pi}{2}-\frac{\pi}{4}\bigr) \right. 
\nonumber \\ 
& & + \left. \frac{z}{2} \left[ r \sqrt{\frac{2}{\pi rz}\,}\cos\bigl(rz-\frac{p\pi}{2}-\frac{\pi}{4}\bigr)
                                  \sqrt{\frac{2}{\pi Rz}\,}\cos\bigl(Rz-\frac{(-p+1)\pi}{2}-\frac{\pi}{4}\bigr) \right. \right.
\nonumber \\
& & \left. \left. ~~~~~~        - R \sqrt{\frac{2}{\pi Rz}\,}\cos\bigl(Rz-\frac{(-p)\pi}{2}-\frac{\pi}{4}\bigr)
                                    \sqrt{\frac{2}{\pi rz}\,}\cos\bigl(rz-\frac{(p+1)\pi}{2}-\frac{\pi}{4}\bigr) \right] \right.
\nonumber \\
& & \left. ~~~~~        +2p\sqrt{\frac{2}{\pi Rz}\,}\cos\bigl(Rz-\frac{(-p)\pi}{2}-\frac{\pi}{4}\bigr)\sqrt{\frac{2}{\pi rz}\,}\cos\big(rz-\frac{p\pi}{2}-\frac{\pi}{4}\bigr) \right\}
\nonumber \\
&\hspace{-1.2cm}=& \hspace{-0.9cm}~\frac{1}{\pi}\lim_{z\to\infty} \left\{ \sqrt{\frac{R}{r}\,} \cos\bigl( Rz-\frac{p\pi}{2}-\frac{3\pi}{4}\bigr)\cos\bigl( rz+\frac{p\pi}{2}-\frac{\pi}{4}\bigr)  
-\sqrt{\frac{r}{R}\,} \cos\bigl(Rz-\frac{p\pi}{2}-\frac{\pi}{4}\bigr)\cos\bigl(rz+\frac{p\pi}{2}-\frac{3\pi}{4}\bigr)  \right.
\nonumber \\
& & \left. +\sqrt{\frac{r}{R}\,}  \cos\bigl(rz -\frac{p\pi}{2}-\frac{\pi}{4}\bigr) \cos\bigl(Rz +\frac{p\pi}{2}-\frac{3\pi}{4}\bigr) 
 -\sqrt{\frac{R}{r}\,} \cos\bigl(Rz+\frac{p\pi}{2}-\frac{\pi}{4}\bigr)\cos\bigl(rz-\frac{p\pi}{2}-\frac{3\pi}{4}\bigr) + {\rm O}\bigl(\frac{1}{z}\bigr) \right\}
\nonumber \\
&\hspace{-1.2cm}=& \hspace{-0.9cm}~\frac{1}{2\pi}\lim_{z\to\infty} \left\{ \sqrt{\frac{R}{r}\,} \left[ \cos\bigl((R-r)z-p\pi-\frac{\pi}{2}\bigr)+\cos\bigl((R+r)z-{\pi}\bigr)\right] 
- \sqrt{\frac{r}{R}\,}\left[ \cos\bigl((R-r)z-p\pi+\frac{\pi}{2}\bigr)+\cos\bigl((R+r)z-\pi\bigr)\right] \right.
\nonumber \\
& & \left. +\sqrt{\frac{r}{R}\,} \left[ \cos\bigl((r-R)z-p\pi+\frac{\pi}{2}\bigr)+\cos\bigl((R+r)z-\pi\bigr)\right] 
- \sqrt{\frac{R}{r}\,} \left[ \cos\bigl((R-r)z+p\pi+\frac{\pi}{2}\bigr) +\cos\bigl((R+r)z-\pi\bigr)\right] \right\}
\nonumber \\
&\hspace{-1.2cm}=& \hspace{-0.9cm}~\frac{1}{2\pi}\lim_{z\to\infty} \left\{ \sqrt{\frac{R}{r}\,} 
\left[ \cos\bigl((R-r)z-p\pi -\frac{\pi}{2}\bigr)-\cos\big((R-r)z+p\pi+\frac{\pi}{2}\bigr) \right] \right. 
\nonumber \\
& & \left. ~~~~~-\sqrt{\frac{r}{R}\,} \left[ \cos\bigl((R-r)z-p\pi+\frac{\pi}{2}\bigr) - \cos\bigl((R-r)z+p\pi-\frac{\pi}{2}\bigr)\right]  \right\}
\nonumber \\
&\hspace{-1.2cm}=& \hspace{-0.9cm}~\frac{1}{2\pi}\lim_{z\to\infty} \left\{ \sin\bigl( (R-r)z\bigr) 
\left[ 2 \sqrt{\frac{R}{r}\,} \sin\bigl( p\pi+\frac{\pi}{2}\bigr) - 2\sqrt{\frac{r}{R}\,}\sin\bigl( p\pi-\frac{\pi}{2}\bigr) \right] \right\}
\nonumber \\
&\hspace{-1.2cm}=& \hspace{-0.9cm}~\frac{\cos p\pi}{\pi} \left[  \sqrt{\frac{R}{r}\,}+ \sqrt{\frac{r}{R}\,}\right] \lim_{z\to\infty} \sin\bigl( (R-r)z\bigr)
\nonumber
\EEA}
We collect these results, insert into ($\#$) and have the statement
\BEA
\lefteqn{ \demi \int_0^{\infty} \!\D u\; u \biggl( J_p(Ru) J_{-p}(ru) +J_{-p}(Ru) J_p(ru) \biggr)} \nonumber \\
&=& \frac{\cos p\pi}{\pi} \left[  \sqrt{\frac{R}{r}\,}+ \sqrt{\frac{r}{R}\,}\,\right] \lim_{z\to\infty} \frac{ \sin\bigl( (R-r)z\bigr)}{(R+r)(R-r)}
+\frac{\sin p\pi}{\pi} \frac{\bigl(R/r\bigr)^p - \bigl(r/R\bigr)^p}{R^2-r^2} 
\nonumber \\
&=& \cos p\pi \frac{\sqrt{R/r\,}  +\sqrt{r/R\,}}{R +r} \underbrace{~\lim_{z\to\infty} \frac{1}{\pi} \frac{\sin\bigl( (R-r)z\bigr)}{R-r}~}_{\to ~\delta(R-r)} 
\:+\,\frac{\sin p\pi}{\pi} \frac{\bigl(R/r\bigr)^p - \bigl(r/R\bigr)^p}{R^2-r^2} 
\nonumber
\EEA
Herein, in the last line, we recognise in the second factor in the first summand a representation of the Dirac distribution $\delta$ in the limit $z\to\infty$\
\cite[p. 38, eq. (4)]{Gelf64}. Since a non-vanishing 
contribution in this summand only arises for $R=r$, the first factor becomes $R^{-1}\cos p\pi$. 
This proves the formal assertion (\ref{gl.B3.a}) which is identical to the rigorous statement (\ref{gl.B2.d}). 

\noindent \underline{Step 2.} Having reproduced the known result (\ref{gl.B2.d}) by our heuristic technique, we now use the same method, built on (*), to discuss the anti-symmetric integral
\BEA
\lefteqn{ \demi \int_0^{\infty} \!\D u\: u \biggl( J_p(uR) J_{-p}(ur) - J_{-p}(uR) J_p(ur) \biggr) \bigl( R^2 - r^2\bigr) }
\nonumber \\
&=& \lim_{z\to\infty} \left\{ \frac{z}{2} \biggl( R J_{p+1}(Rz) J_{-p}(rz) - r J_{p}(Rz) J_{-p}(rz) \biggr) -p J_{p}(Rz) J_{-p}(rz) \right. 
\nonumber \\
& & \left. ~~~~-  \frac{z}{2} \biggl( R J_{-p+1}(Rz) J_{p}(rz) - r J_{-p}(Rz) J_{p+1}(rz) \biggr) -p J_{-p}(Rz) J_{p}(rz) \right\} 
\nonumber \\
& & -\lim_{z\to 0} \left\{ \frac{z}{2} \biggl( R J_{p+1}(Rz) J_{-p}(rz) - r J_{p}(Rz) J_{-p}(rz) \biggr) -p J_p(Rz) J_{-p}(rz) \right. 
\nonumber \\
& & \left. ~~~~-  \frac{z}{2} \biggl( R J_{-p+1}(Rz) J_{p}(rz) - r J_{-p}(Rz) J_{p+1}(rz) \biggr) -p J_{-p}(Rz) J_{p}(rz) \right\} 
\nonumber 
\EEA 
such that the sought limit decomposes into  ${\cal L}=T_1 - T_2$ which are found separately, as before. The second term reads,  with the standard low-order expansion 
\BD
T_2 = \lim_{z\to 0} \left\{ {\rm O}(z^2) -\frac{p}{\Gamma(1+p)\Gamma(1-p)} \left[  \sqrt{\frac{R}{r}\,}+ \sqrt{\frac{r}{R}\,}\right] z^0 \right\} 
= -\frac{\sin p\pi}{\pi} \left[  \sqrt{\frac{R}{r}\,}+ \sqrt{\frac{r}{R}\,}\,\right]
\ED
and first first term reads, in analogy with the previous asymptotic and trigonometric calculations
{\small\BEA 
\lefteqn{ \hspace{-1.2cm}T_1 = \frac{1}{2\pi}\lim_{z\to\infty} \left\{ \sqrt{\frac{R}{r}\,} \left[ \cos\bigl((R-r)z-p\pi-\frac{\pi}{2}\bigr)+\cos\bigl((R+r)z-{\pi}\bigr)\right] 
- \sqrt{\frac{r}{R}\,}\left[ \cos\bigl((R-r)z-p\pi+\frac{\pi}{2}\bigr)+\cos\bigl((R+r)z-\pi\bigr)\right] \right.}
\nonumber \\
& & \left. -\sqrt{\frac{r}{R}\,} \left[ \cos\bigl((r-R)z-p\pi+\frac{\pi}{2}\bigr)+\cos\bigl((R+r)z-\pi\bigr)\right] 
+ \sqrt{\frac{R}{r}\,} \left[ \cos\bigl((R-r)z+p\pi+\frac{\pi}{2}\bigr) +\cos\bigl((R+r)z-\pi\bigr)\right] \right\}
\nonumber \\
&\hspace{-1.2cm}=& \hspace{-0.9cm}~\frac{1}{\pi}\lim_{z\to\infty} \left\{ \sqrt{\frac{R}{r}\,} 
\left[ \cos\bigl((R-r)z\bigr) \sin p\pi - \cos\bigl((R+r)z\bigr)\right] -\sqrt{\frac{r}{R}\,} \left[ \cos\bigl((R-r)z\bigr) (-\sin p\pi) -\cos\bigl((R+r)\bigr) \right]  \right\}
\nonumber \\
&\hspace{-1.2cm}=& \hspace{-0.9cm}~\frac{1}{\pi} \lim_{z\to\infty} \left\{ \sin p\pi \left[ \sqrt{\frac{R}{r}\,}+\sqrt{\frac{r}{R}\,}\right]\cos\bigl( (R+r)z\bigr) 
-\left[ \sqrt{\frac{R}{r}\,}-\sqrt{\frac{r}{R}\,}\,\right]\cos\bigl( (R+r)z\bigr) \right\}
\nonumber
\EEA}
For the interpretation of these improper limits, consider the following transformation
\BEA
\lefteqn{\frac{\cos xz}{\pi x} = \frac{\cos xz \,-1 +1}{\pi x} = \frac{1}{\pi x} - \frac{1-\cos xz}{\pi x} }\nonumber \\
&=& \frac{1}{\pi x} -\frac{1}{\pi} \int_0^z \!\D y\: \sin yx \:=\: \frac{1}{\pi x} - \frac{1}{2\pi \II} \int_0^z \!\D y \: \biggl( e^{\II x y} - e^{-\II x y} \biggr) 
\nonumber \\
&=& \frac{1}{\pi x} - \frac{1}{2\pi\II} \left[ \int_0^z \!\D y\: e^{\II x y} - \int_{-z}^0 \!\D y\: e^{\II x y} \right] 
\nonumber \\
&=& \frac{1}{\pi x} - \frac{1}{2\pi\II} \int_{-z}^z \!\D y\: e^{\II x y}\,  \sgn(y) \;\; ; \;\; \mbox{\rm\small ~with~} 
\sgn(y) := \left\{ \begin{array}{ll} +1 & \mbox{\rm\small ~~;~ if $y>0$} \\
                                     -1 & \mbox{\rm\small ~~;~ if $y<0$}
                   \end{array} \right.                                                                                                        
\nonumber 
\EEA  
Now, in the limit $z\to\infty$, this can be recognised as the Fourier transform $\mathcal{F}$ of the distribution $\sgn(y)$ (we use the notation and conventions of \cite{Gelf64}) and
\BD
\lim_{z\to\infty} \frac{\cos xz}{\pi x} = \frac{1}{\pi x} -  \frac{1}{2\pi\II} \mathcal{F}\bigl(\sgn(y)\bigr)(x) =   \frac{1}{\pi}\frac{1}{x} -  \frac{1}{2\pi\II} 2\II\frac{1}{x}  = 0 \tag{**} 
\ED
because of the distributional identity $\mathcal{F}\bigl(\sgn(y)\bigr)(x)=\frac{2\II}{x}$ \cite[p. 173, eq. (15) \& p. 174, eq. (17)]{Gelf64}. 

At long last, we can combine all this and find
\BEA
\lefteqn{ \demi \int_0^{\infty} \!\D u\: u \biggl( J_p(uR) J_{-p}(ur) - J_{-p}(uR) J_p(ur) \biggr)  }
\nonumber \\
&=& \sin p\pi \frac{\sqrt{R/r\,}+\sqrt{r/R\,}}{R+r} \underbrace{~\lim_{z\to\infty} \frac{\cos (R-r)z}{\pi (R-r)}~}_{\to ~0} 
-  \frac{\sqrt{R/r\,}-\sqrt{r/R\,}}{R-r} \underbrace{~\lim_{z\to\infty} \frac{\cos (R+r)z}{\pi (R+r)}~}_{\to ~0} 
\nonumber \\
&& + \frac{\sin p\pi}{\pi} \frac{\bigl(R/r\bigr)^p + \bigl(r/R\bigr)^p}{R^2 -r^2}
\nonumber
\EEA
where the two improper limits vanish because of (**). This proves formally also the second assertion (\ref{gl.B3.b}). \hfill q.e.d.

Notice the structural similarity with the results derived on the finite interval $[0,1]$ in appendix~A. 

Our arguments were based on formal asymptotic expansions, to leading order and on borrowing some identities from the theory of distributions \cite{Schw1950,Gelf64}. 
We leave as an open problem the construction of a mathematically rigorous proof of the identities (\ref{gl.B3.b},\ref{gl.B4}). 

\newpage
\appsection{C}{On Hilbert and Mellin transforms} 

To make this work more self-contained, we provide background on the Hilbert transform on the positive real line $\mathbb{R}_+$  \cite{Pave94,Blas23}
(for brevity we shall refer to it simply as `Hilbert transform') and the Mellin transform \cite{Flaj95}. 
In this appendix, all integrals are understood as Cauchy principal values. 


\noindent
{\bf Lemma C1.} \cite{Pap99} {\it If $0<u<1$, one has}
\BEQ \label{gl.C1}
I := \int_0^{\infty}  \!\D t\: \frac{t^{u-1}}{t-1} = - \pi \cot\pi u
\EEQ

\begin{figure}[tb]
\includegraphics[width=0.50\hsize]{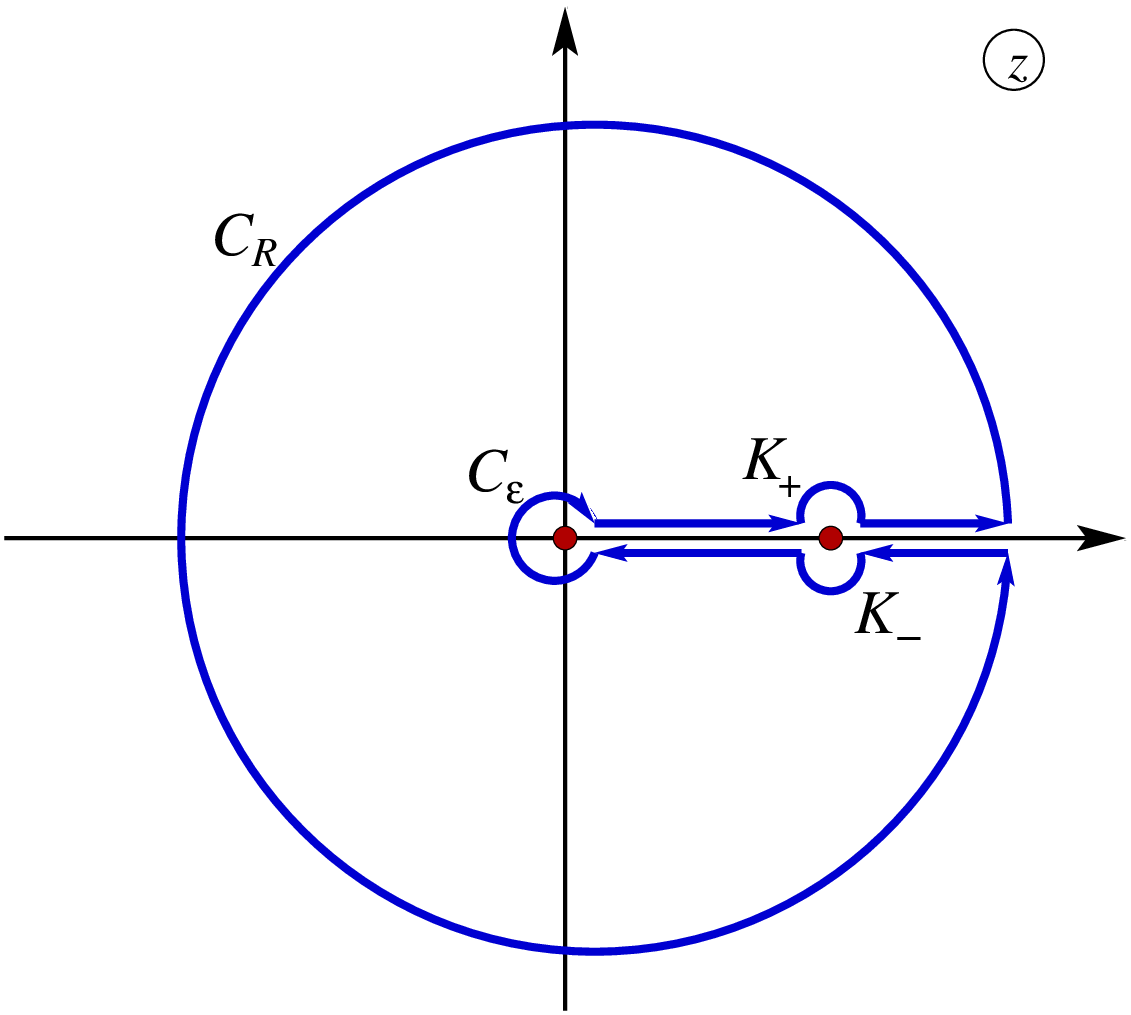}  
\caption[figC1]{Integration contour $\mathscr{C}$ needed in the proof of (\ref{gl.C1}).
\label{figC1} }
\end{figure}

\noindent
{\bf Proof.} We follow closely \cite[pp. 218-220]{Pap99}. Consider the contour integral
\BD
J := \oint_{\mathscr{C}} \!\D z\: \frac{z^{u-1}}{z-1} = 
\left(\int_{\vep}^{1-\vep'} +\int_{K_+} + \int_{1+\vep'}^R +  \int_{C_R} + \int_{R}^{1+\vep'} + \int_{K_-} + \int_{1-\vep'}^{\vep} + \int_{C_{\vep}}\right) 
\!\D z\: \frac{z^{u-1}}{z-1} 
\ED
where the integration contour $\mathscr{C}$ is sketched in figure~\ref{figC1}. 
Since the two singularities (red points in figure~\ref{figC1}) are not in the interior of
$\mathscr{C}$, the contour integral $J=0$. Here, the segments $C_R$, $C_{\vep}$ 
are circles around zero, of radius $R,\vep$ respectively, and $K_{\pm}$ are half-circles around $1$ of radius
$\vep'$. The limits $\vep\to 0$, $\vep'\to 0$ and $R\to\infty$ are to taken at an appropriate moment. 
Since the integral converges in the $\vep\to 0$ and $R\to\infty$ limits, the main focus
must be on how to take $\vep'\to 0$ correctly. 

First, since $0<u<1$, one can estimate
\BD
\left| \int_{C_R}\!\D z\: \frac{z^{u-1}}{z-1} \right| \leq \int_{C_R}\!\bigl|\D z\bigr|\: \left| \frac{z^{u-1}}{z-1} \right| \sim R \frac{R^{u-1}}{R} \to 0
\;\; ,  \;\;
\left| \int_{C_{\vep}}\!\D z\: \frac{z^{u-1}}{z-1} \right| \leq \int_{C_{\vep}}\!\bigl|\D z\bigr|\: \left| \frac{z^{u-1}}{z-1} \right| \sim \vep^u \to 0
\ED
when $R\to\infty$ and $\vep\to 0$, respectively. Second, with the choice $z=1+\vep' e^{\II\theta}$, we have
\BD
\int_{K_+}\!\!\D z\: \frac{z^{u-1}}{z-1} = \int_{\pi}^{0} \!\D\theta\: \II\vep'\,e^{\II\theta} \frac{\bigl(1+\vep' e^{\II\theta}\bigr)^{u-1}}{\vep' e^{\II\theta}} 
                                         = -\II \int_0^{\pi} \!\D\theta\: \sum_{k=0}^{\infty} \binom{u-1}{k} {\vep'}^k e^{\II k \theta} = -\II \pi + {\rm O}(\vep')
\ED
and in the other case, we let $z=\bigl(1-\vep' e^{\II\theta}\bigr)e^{2\pi \II}$ and have
\BEA
\int_{K_-}\!\!\D z\: \frac{z^{u-1}}{z-1} &=& -\int_{0}^{\pi} \!\D\theta\: \bigl(-\II\vep'\bigr)\,e^{\II\theta} 
                                             \frac{\bigl(1+\vep' e^{\II\theta}\bigr)^{u-1}e^{2\pi\II(u-1)}}{-\vep' e^{\II\theta}} 
\nonumber \\
&=& -\II\, e^{2\pi\II (u-1)} \int_0^{\pi} \!\D\theta\: \sum_{k=0}^{\infty} \binom{u-1}{k} \bigl(-{\vep'}\bigr)^k e^{\II k \theta} \:=\: -\II \pi e^{2\pi\II (u-1)} + {\rm O}(\vep')
\nonumber 
\EEA
where the extra sign arises from the opposite orientation of the contour. Third, the contributions from the straight lines can now be written as
\BD
I_R := \lim_{\vep,\vep'\to 0} \left(  \int_{\vep}^{1-\vep'} + \int_{1+\vep'}^R \right) \D z\: \frac{z^{u-1}}{z-1}
\ED
and 
\BD 
\lim_{\vep,\vep'\to 0}  \left( \int_{R}^{1+\vep'} + \int_{1-\vep'}^{\vep} \right) \D z\: \frac{\bigl(e^{2\pi \II}z\bigr)^{u-1}}{z-1} = - e^{2\pi \II\bigl(u-1\bigr)} I_R
\ED
where the extra sign comes from the opposite orientation for the integration contour 
and the change $z\mapsto z e^{2\pi \II}$ is necessary in order to remain on the same sheet. Forth, taking all this
together, we have (anticipating that $R\to \infty$ will be taken shortly) 
\BD 
J = 0 = \left( 1 - e^{2\pi \II\bigl(u-1\bigr)} \right) I_R-\II \pi \left( 1 + e^{2\pi \II \bigl(u-1\bigr)} \right)
\ED
which finally allows to conclude
\BD 
I = \lim_{R\to\infty} I_R = \II \pi\, \frac{1+e^{2\pi \II u}}{1-e^{2\pi \II u}} =\II \pi\, \frac{e^{-\pi \II u} + e^{\pi \II u}}{e^{-\pi \II u} - e^{\pi \II u}} 
= -\pi\, \frac{\cos \pi u}{\sin \pi u} = -\pi \cot\pi u
\ED
as claimed. \hfill q.e.d. 

Once established in the strip $0<u<1$, one may use (\ref{gl.C1}) to define the integral $I=I(u)$ for other values of $u$ via analytic continuation. 

On the half-line $\mathbb{R}_+$, the {\em Hilbert transform} $f(x)$ of a function $g(y)$ is defined as
\BEQ \label{gl.C2}
f(x) = \mathscr{H}\bigl(g(y)\bigr)(x) := \frac{1}{\pi}\int_0^{\infty} \!\D y\: \frac{g(y)}{y-x} \;\; , \;\;
g(y) = -\frac{1}{\pi} \int_0^{\infty} \!\D x\: \sqrt{ \frac{x}{y}\, } \,\frac{f(x)}{x-y}
\EEQ
where the integrals are Cauchy principal values. We implicitly admit that $g(y)$ satisfies sufficient conditions such that its Hilbert transform $f(x)$ exists. 
The inverse Hilbert transformation also included in (\ref{gl.C2}) was derived in \cite{Pave94} 
(the square root is distinct with respect to an inverse Hilbert transform on the entire real line $\mathbb{R}$). 

In addition, we need the {\em Mellin transform} of a function $f(x)$, defined as \cite{Flaj95} 
\BEQ \label{gl.C3}
\mel{f}(u) = \mathscr{M}\bigl(f(x)\bigr)(u) := \int_0^{\infty} \!\D x\: x^{u-1} f(x) \;\; , \;\;
f(x) = \frac{1}{2\pi \II} \int_{c-\II\infty}^{c+\II\infty} \!\D u \: x^{-u} \mel{f}(u)
\EEQ
For the existence of a Mellin transform of a function $f(x)$, a consideration of its {\it fundamental strip} $\langle a, b\rangle$ with $a<b$ is paramount. Herein
\BEA
a  &:=& \mbox{\rm inf}\;\left( A \in\mathbb{R} \left| \mbox{\rm $f(x) x^{A-1}$ is integrable over $(0,1]$}       \right. \right) \nonumber \\
b  &:=& \mbox{\rm sup}  \left( B \in\mathbb{R} \left| \mbox{\rm $f(x) x^{B-1}$ is integrable over $[1,\infty)$}  \right. \right) \nonumber 
\EEA
In practise, an useful shortcut goes as follows, which appeals to the asymptotic behaviour of $f(x)$. 
If $f(x)\stackrel{x\to 0}{\sim} x^{-\alpha}$ and $f(x)\stackrel{x\to\infty}{\sim} 
x^{-\beta}$ with $\alpha<\beta$, then the fundamental strip $\langle a, b\rangle$ is a superset of the 
strip $\left\{ u = \sigma +\II \tau \in\mathbb{C} \bigl|  \tau\in\mathbb{R} 
\mbox{\rm  ~and~} \alpha<\sigma<\beta\bigr.\right\}$. On the fundamental strip, 
$\mel{f}(u)$ exists and is holomorphic \cite{Flaj95}. A monomial $f(x)=x^a$ does not admit a Mellin transform. The constant $c$ in (\ref{gl.C3}) must be in the fundamental strip. 

\noindent 
{\bf Lemma C.2} {\it If both the Hilbert and Mellin transforms of a function $f(x)$ exist, as well as the Mellin transform of $\mathscr{H}(f)(x)$, and $0<u<1$, then formally}
\BEQ \label{gl.C4} 
\mathscr{M}\left(\mathscr{H}\bigl(f(x)\bigr)\right)(u) = - \cot\pi u \:\mathscr{M}\bigl(f(x)\bigr)(u)
\EEQ

\noindent
{\bf Proof.} Under the hypotheses made on the existence of the objects in question, this is merely a formal calculation
\BEA
\lefteqn{ \mathscr{M}\left(\mathscr{H}\bigl(f(x)\bigr)\right)(u) = \int_0^{\infty} \!\D x\: x^{u-1} \,\frac{1}{\pi} \int_0^{\infty} \!\D y\: \frac{f(y)}{x-y} 
\:=\: \frac{1}{\pi} \int_0^{\infty} \!\D y\: f(y) \int_{0}^{\infty} \!\D x\: \frac{x^{u-1}}{x-y} }
\nonumber \\
&=& \frac{1}{\pi} \int_0^{\infty} \!\D y\: f(y) \int_{0}^{\infty} \!\D t\: y\, \frac{\bigl(yt\bigr)^{u-1}}{yt -y}
\:=\: \frac{1}{\pi} \int_0^{\infty} \!\D y\: y^{u-1}\, f(y) \int_{0}^{\infty} \!\D t\: \frac{t^{u-1}}{t-1}
\nonumber
\EEA
and the assertion follows from the identity (\ref{gl.C1}). \hfill q.e.d. 

As an application, one may treat integral equations of the type $\mathscr{H}(f)(x)=g(x)$ \cite{Blas23} 
where $g(x)$ is given and $f(x)$ is sought. A Mellin transformation of this gives
\BD 
\mathscr{M}\bigl(\mathscr{H}(f)\bigr)(u)= -\cot\pi u\, \mathscr{M}(f)(u) = \mathscr{M}(g)(u)
\ED 
hence formally $\mathscr{M}(f)(u)=-\tan\pi u\,  \mathscr{M}(g)(u)$. See \cite{Blas23} for more details. 
It remains to carry out efficiently the inverse Mellin transformation, in order to really find $f(x)$. 

\noindent
{\bf Lemma C.3} {\it If the function $f(x)$ is such that all integrals and Mellin transformations which follow exist, and $0<u-p<1$, one has formally}
\BEQ \label{gl.C5} 
\mathscr{M}\left( \int_0^{\infty} \!\D x \left( \frac{x}{y} \right)^p \frac{f(x)}{x-y} \right)(u) = \pi \cot\bigl( \pi(u-p)\bigr) \mathscr{M}\bigl(f(x)\bigr)(u)
\EEQ

\noindent
{\bf Proof:} This follows from a formal calculation, as follows:
\BEA
\lefteqn{ \mathscr{M}\left( \int_0^{\infty} \!\D x \left( \frac{x}{y} \right)^p \frac{f(x)}{x-y} \right)(u) 
\hspace{2.35cm}= \int_0^{\infty} \!\D y\: y^{u-1} \int_0^{\infty} \!\D x \left( \frac{x}{y} \right)^p \frac{f(x)}{x-y} }\nonumber \\
&=& \int_0^{\infty} \!\D x\: f(x) \int_0^{\infty} \!\D y\: y^{u-1} \left( \frac{x}{y} \right)^p \frac{1}{x-y} 
\:=\: \int_0^{\infty} \!\D x\: f(x) \int_0^{\infty} \!\D t\: x \bigl(t x\bigr)^{u-1} \left( \frac{x}{t x} \right)^p \frac{1}{x-tx} \nonumber \\
&=& \int_0^{\infty} \!\D x\: x^{u-1}\, f(x) \int_0^{\infty} \!\D t\:  \frac{t^{u-p-1}}{1-t} 
\hspace{1.05cm}\:=\: \pi \mathscr{M}\bigl(f(x)\bigr)(u) \cdot \cot\bigl( \pi(u-p)\bigr) \nonumber
\EEA
as claimed, with the help of Lemma C.1. \hfill q.e.d. 

Of course, once eqs.~(\ref{gl.C4},\ref{gl.C5}) are obtained in some strip, one may extend these via analytic continuation. They are required in the main text. 

A few simple rules, besides the obvious properties of linearity and scaling, include the following \cite{Flaj95}
\begin{subequations} \label{gl:Melprops}
\begin{align}
\mathscr{M}\left( f\bigl(1/x) \right)(u) &= - \mathscr{M}\bigl(f(x)\bigr)(-u) \:=\: -\mel{f}(-u) \\
\mathscr{M}\left( f\bigl(x\bigr) \ln x\right)(u) &= \frac{\D}{\D u}\mathscr{M}\bigl(f(x)\bigr)(u) ~ \:=\: \frac{\D}{\D u}\mel{f}(u) \\
\mathscr{M}\left( x \frac{\D}{\D x} f(x)\right)(u) &= - u\mathscr{M}\bigl(f(x)\bigr)(u) \:=\: -u\mel{f}(u) \label{gl:Melprops3}
\end{align}
\end{subequations}
where in (\ref{gl:Melprops3}) $f(x)$ is assumed continuous and piecewise differentiable. 

The asymptotic form of $f(x)$ as $x\to 0$ ($x\to \infty$) constrains the behaviour of $\mel{f}(u)$ at the
left (right) boundary of its fundamental strip $\langle a, b\rangle$ and inversely. 
For a meromorphic function $\phi(u)$ with a p\^ole of order $r>0$, 
its {\em singular element} at $u_0$ is the sum of the singular terms in its Laurent expansion at $u_0$ and its {\em singular expansion} is the formal sum of its
singular elements at all points $u_0$. If $E$ is such a singular expansion, one writes $\phi(u)\asymp E$. 
For example $\Gamma(u)\asymp\sum_{\ell=0}^{\infty} \frac{(-1)^{\ell}}{\ell !} \frac{1}{u+\ell}$ \cite{Abra65}. 

The relationship with p\^oles of the Mellin transform is expressed in the following two theorems. 
For simplicity, it is admitted throughout that $f(x)$ is continuous on $(0,\infty)$, 
and has a non-empty fundamental strip $\langle a, b\rangle$ where the Mellin transform  $\mel{f}(u)$ exists. 

\noindent 
{\bf Theorem C.1} (Direct mapping theorem \cite{Flaj95}). {\it (i) If for $x\to 0^+$ one has a finite expansion}
\BEQ \label{gl:C-fasymp}
f(x) = \sum_{\xi,k} \mathfrak{c}_{\xi,k}\, x^{\xi} \ln^k x + {\rm O}\bigl(x^{c}\bigr) \;\; ; \;\; -c < -\xi \leq a \;\; , \;\; k\in\mathbb{N} 
\EEQ
{\it with known coefficients $\mathfrak{c}_{\xi,k}$, then $\mel{f}(u)$ can be continued to a meromorphic function in the strip $\langle -c,b\rangle$ with the singular expansion}
\begin{subequations}
\BEQ
\mel{f}(u) \asymp \sum_{\xi,k}\, \mathfrak{c}_{\xi,k} \frac{(-1)^k\, k!}{\bigl(u+\xi\bigr)^{k+1}} \;\; ; \;\; u\in\langle -c,b\rangle \;\; , \;\; k\in\mathbb{N}
\EEQ
{\it (ii) Similarly, if $f(x)$ admits for $x\to\infty$ an asymptotic expansion of the form (\ref{gl:C-fasymp}) with $b\leq \xi<-c$, $\mel{f}(u)$ can be continued to 
a meromorphic function in the strip $\langle a, -c\rangle$ and}
\BEQ
\mel{f}(u) \asymp- \sum_{\xi,k} \mathfrak{c}_{\xi,k}\, \frac{(-1)^k\, k!}{\bigl(u+\xi\bigr)^{k+1}} \;\; ; \;\; u\in\langle a, -c\rangle \;\; , \;\; k\in\mathbb{N}
\EEQ
\end{subequations}

\noindent
{\bf Remark:} subtracting from $f(x)$ a truncated form of its asymptotic expansions at $0$ or $\infty$ merely shifts the fundamental strip but does not
change the formal singular expansion of $\mel{f}(u)$. 

\noindent 
{\bf Theorem C.2} (Converse mapping theorem \cite{Flaj95}). {\it (i) If $\mel{f}(u)$ has a meromorphic continuation to the strip $\langle c,b\rangle$ with $c<a$ 
with a finite number of p\^oles, is analytic for $\Re u=c$ and for $\eta\in(a,b)$ obeys the growth condition, for $c\leq \Re u\leq \eta$}
\BEQ \label{gl:C-growth}
\mel{f}(u) = {\rm O}\bigl(|u|^{-r}\bigr) \;\; , \;\; \mbox{\it as $|u|\to\infty$ and with $r>1$}
\EEQ
{\it then its singular expansion in the strip $u\in \langle c,a\rangle$ (to the left of the fundamental strip)}
\BEQ \label{gl:C-melsing}
\mel{f}(u) \asymp \sum_{\xi,k} \mathfrak{d}_{\xi,k}\, \bigl( u - \xi\bigr)^{-k} \;\; ; \;\; \mbox{\it\small ~~where  $k\geq 1$~~} 
\EEQ
{\it with known coefficients $\mathfrak{d}_{\xi,k}$ implies the asymptotic expansion for $x\to 0^+$}
\begin{subequations}
\BEQ \label{gl:C-melx0}
f(x) = \sum_{\xi,k} \mathfrak{d}_{\xi,k-1} \left(\frac{(-1)^{k-1}}{(k-1)!}\, x^{-\xi} \ln^{k-1} x \right) + {\rm O}\bigl(x^{-c}\bigr)
\EEQ
{\it (ii) Similarly, if $\mel{f}(u)$ has a meromorphic continuation to the strip $\langle a, c\rangle$ with $c>b$, has the same properties as before but satisfies the
growth condition (\ref{gl:C-growth}) as well as the singular expansion (\ref{gl:C-melsing}) in the strip $\langle \eta,c\rangle$, one has asymptotically for $x\to\infty$}
\BEQ\label{gl:C-melxinf}
f(x) = -\sum_{\xi,k} \mathfrak{d}_{\xi,k-1} \left(\frac{(-1)^{k-1}}{(k-1)!}\, x^{-\xi} \ln^{k-1} x \right) + {\rm O}\bigl(x^{-c}\bigr)
\EEQ
\end{subequations}

\noindent
{\bf Remark.} This result allows to identify the asymptotic behaviour of a function $f(x)$ if the singular expansion (\ref{gl:C-melsing}) of its Mellin transform $\mel{f}(u)$ is known. 

\noindent
{\bf Example.} Consider the following function (with $\nu>0$), analytic in its fundamental strip $\langle 0,\nu\rangle$, together with its singular expansion to the left of $\Re u=0$ \cite{Flaj95}
\BD
\phi(u) = \frac{\Gamma(u)\Gamma(\nu-u)}{\Gamma(\nu)} \asymp \sum_{\ell=0}^{\infty} \frac{(-1)^{\ell}}{\ell !} \frac{\Gamma(\nu+\ell)}{\Gamma(\nu)} \frac{1}{u+\ell} 
\;\; ; \;\; \Re u <0 
\ED
which only contains simple p\^oles, all on the non-positive real axis ($k=1$ in (\ref{gl:C-melsing})). 
The properties of the Gamma function guarantee a sufficiently fast decrease on the strip such that Theorem C.2 is applicable. Since the singular expansion is to the left of the 
fundamental strip, this gives the small-$x$ behaviour of the original function. 
Eq.~(\ref{gl:C-melx0}) with $k=1$ implies
\BD
f(x) = \frac{1}{2\pi\II} \int_{\nu/2-\II\infty}^{\nu/2+\II\infty} \!\!\D u\: x^{-u}\,\phi(u) 
= \sum_{\ell=0}^L \frac{(-1)^{\ell}}{\ell !} \frac{\Gamma(\ell+\nu)}{\Gamma(\nu)}  x^{\ell} + {\rm O}\bigl( x^{L+1/2}\bigr)
= \bigl(1+x\bigr)^{-\nu} \tag{*}
\ED
if one lets $L\to\infty$, up to terms vanishing more fast than any power (see \cite{Flaj95} for details). 
Alternatively, we can write the singular expansion to the right of $\Re u = \nu>0$, which reads
\BD 
\phi(u) = \frac{\Gamma(u)\Gamma(\nu-u)}{\Gamma(\nu)} \asymp - \sum_{\ell=0}^{\infty} \frac{\Gamma(\nu+\ell)}{\Gamma(\nu)} \frac{(-1)^{\ell}}{\ell !} \frac{1}{u-\bigl(\nu+\ell\bigr)}
\;\; ; \;\; \Re u > \nu 
\ED
Since this is on the right of the fundamental strip, this produces the large-argument asymptotics. Now, eq.~(\ref{gl:C-melxinf}) with $k=1$ implies for the original function
\BD
f(x) = - (-1) \sum_{\ell=0}^{\infty} \frac{\Gamma(\nu+\ell)}{\Gamma(\nu)} \frac{(-1)^{\ell}}{\ell !} x^{-\nu-\ell} \tag{**}
\ED
In order to check whether this reproduces indeed the form $(1+x)^{-\nu}$ expected from (*), we expand via the binomial theorem
\BD
\bigl(1+x\bigr)^{-\nu} = x^{-\nu} \sum_{\ell=0}^{\infty} \frac{1}{\ell !} \frac{\Gamma(1-\nu)}{\Gamma(1-\nu-\ell)}\, x^{-\ell} 
= \sum_{\ell=0}^{\infty} \frac{1}{\ell !} \underbrace{\frac{\Gamma(1-\nu)\Gamma(\nu)}{\Gamma(1-\nu-\ell)\Gamma(\nu+\ell)}}_{=(-1)^{\ell}} \frac{\Gamma(\nu+\ell)}{\Gamma(\nu)}\,x^{-\nu-\ell} 
\ED
and find agreement with (**) where we used \cite[(6.1.21),(6.1.17)]{Abra65}. 
Hence both expansions show that $f(x), \phi(u)=\mel{f}(u)$ are a Mellin transform pair. 
Of course, this can also be verified directly, e.g. via a Beta-function identity \cite{Abra65}.

See \cite{Flaj95} for further examples and illustrations. 

\appsection{D}{On some mathematical identities} 

\noindent 
{\bf 1.} We prove the two identities (\ref{gl:4.15}) on the functions $\mathscr{G}_{\pm}(u)$. This uses the known scaling function $F(1;u)$ of the single-time correlator. 

First, we recall special cases of the identities \cite[(2.12.2.2)]{Prud2} and \cite[(1.2.1.1)]{Prud2}, namely 
\begin{subequations} \label{gl:D.1}
\begin{align}
& \int_0^{\infty} \!\D y\: y^{2u-p-1} J_{-p}(2y) = \demi \frac{\Gamma(u-p)}{\Gamma(1-u)} \label{gl:D.1a} \\
& \int_0^{\infty} \!\D v\: v^{-u} \Gamma\left(p,\frac{v}{(2+\theta)^2} \right) 
= \bigl(2+\theta\bigr)^{2(1-u)} \frac{\Gamma\bigl(p+1-u\bigr)}{1-u} \label{gl:D.1b}
\end{align}
\end{subequations}
(with $0<p<1$ and implicit limits implied by the singularities of the integrals) and then find
\BEA
\mathscr{G}_{-}(u) &=& \int_{0}^{\infty} \!\D\mu\: \mu^{u-1-p/2} \int_0^{\infty} \!\D v\: v^{-p/2} F(1;v) J_{-p}\bigl(2\sqrt{\mu v\,}\,\bigr) 
\nonumber \\
&=& \frac{1}{\Gamma(p)} \int_0^{\infty} \!\D v\: v^{-p/2} \Gamma\left(p,\frac{v}{(2+\theta)^2} \right)  
                        \int_0^{\infty} \!\D w\: v^{-1} \left( \frac{w}{v}\right)^{u-1-p/2} J_{-p}\bigl(2\sqrt{w\,}\,\bigr) 
\nonumber \\
&=& \frac{2}{\Gamma(p)}\int_0^{\infty} \!\D v\: v^{-u} \Gamma\left(p,\frac{v}{(2+\theta)^2} \right)
                       \int_0^{\infty} \!\D y\: y^{2u-p-1} J_{-p}\bigl(2 y\bigr) 
\nonumber \\
&=& \frac{\bigl(2+\theta\bigr)^{2(1-u)}}{\Gamma(p)} \frac{\Gamma(u-p)}{\Gamma(2-u)}\Gamma\left(p+1-u\right)
\label{gl:D.2}
\EEA
and in the last step, we used the identities (\ref{gl:D.1a},\ref{gl:D.1b}). This gives (\ref{gl:4.15a}). 
Second, we need another special case of \cite[(2.12.2.2)]{Prud2}, namely 
\addtocounter{equation}{-2}
\begin{subequations}
\addtocounter{equation}{2}
\begin{align}
& \int_0^{\infty} \!\D y\: y^{2u-p-1} J_{p}(2y) = \demi \frac{\Gamma(u)}{\Gamma(1+p-u)} \label{gl:D.1c}
\end{align}
\end{subequations}
\addtocounter{equation}{1}
in order to find
\BEA
\mathscr{G}_{+}(u) &=& \int_{0}^{\infty} \!\D\mu\: \mu^{u-1-p/2} \int_0^{\infty} \!\D v\: v^{-p/2} F(1;v) J_{+p}\bigl(2\sqrt{\mu v\,}\,\bigr) 
\nonumber \\
&=& \frac{1}{\Gamma(p)} \int_0^{\infty} \!\D v\: v^{-p/2} \Gamma\left(p,\frac{v}{(2+\theta)^2} \right)  
                        \int_0^{\infty} \!\D w\: v^{-1} \left( \frac{w}{v}\right)^{u-1-p/2} J_{p}\bigl(2\sqrt{w\,}\,\bigr) 
\nonumber \\
&=& \frac{2}{\Gamma(p)}\int_0^{\infty} \!\D v\: v^{-u} \Gamma\left(p,\frac{v}{(2+\theta)^2} \right)
                       \int_0^{\infty} \!\D y\: y^{2u-p-1} J_{p}\bigl(2 y\bigr) 
\nonumber \\
&=& -\frac{\bigl(2+\theta\bigr)^{2(1-u)}}{\Gamma(p)} \frac{\Gamma(u-1)}{\Gamma(1+p-u)} \Gamma\left(p+1-u\right) 
\label{gl:D.3} 
\EEA
where in the last step, we used the identities (\ref{gl:D.1b},\ref{gl:D.1c}) and have proven (\ref{gl:4.15b}). 

These two functions are not independent, but are related by the identity 
\BEQ \label{gl:D.4}
\mathscr{G}_-(u)-\bigl( \cos\pi p + \sin \pi p \cot\pi(u-p)\bigr)\mathscr{G}_+(u) =0
\EEQ

\noindent 
{\bf 2.} In order to obtain the two-time auto-correlator for $d_s<2$, we need the Laplace transform of the function $\mathscr{F}_0(\kappa)$.
As a preparation, we recall the identity (\ref{gl:D.1b}) and \cite[(11.4.29)]{Abra65}, for $0<p<1$
\addtocounter{equation}{-4}
\begin{subequations}
\addtocounter{equation}{3}
\begin{align}
\int_0^{\infty} \!\D y\: y^{1-p}\, e^{-\bigl(\sigma/u\bigr) y^2} J_{-p}(2y) = \demi \left( \frac{u}{\sigma}\right)^{1-p} e^{-u/\sigma}  \label{gl:D.1d}
\end{align}
\end{subequations}
\addtocounter{equation}{3}
and then have 
\begin{subequations}
\begin{align}
\lap{\mathscr{F}_0(\kappa)}(\sigma) &= \int_0^{\infty} \!\D\kappa\: e^{-\sigma\kappa} \mathscr{F}_0(\kappa) 
\:=\: \int_0^{\infty} \!\D\kappa\: \kappa^{-p/2}\, e^{-\sigma\kappa} \int_0^{\infty} \!\D u\: F(1;u) u^{-p/2} J_{-p}\bigl(2\sqrt{\kappa u\,}\,\bigr)
\nonumber \\
&= \int_0^{\infty} \!\D u\: u^{-p/2} F(1;u) \int_0^{\infty} \!\D w\: u^{-1} \left( \frac{w}{u}\right)^{-p/2} e^{-\sigma \frac{w}{u}} J_{-p}\bigl(2\sqrt{w\,}\,\bigr)
\nonumber \\
&= 2 \int_0^{\infty} \!\D u\: u^{-1} F(1;u) \int_0^{\infty} \!\D y\: y^{1-p} \exp\left(-\frac{\sigma}{u} y^2\right) J_{-p}(2y) 
\nonumber \\
&= \sigma^{p-1} \int_0^{\infty} \!\D u\: F(1;u) u^{-p} e^{-u/\sigma} 
\:=\: \frac{\sigma^{p-1}}{\Gamma(p)} \int_0^{\infty} \!\D u\: \Gamma\left(p,\frac{u}{(2+\theta)^2} \right) u^{-p}\, e^{-u/\sigma} 
\label{gl:D.5a} \\
&= \frac{\sigma^{p-1}}{\Gamma(p)} \sum_{n=0}^{\infty} \frac{(-1)^n}{n!} \frac{1}{\sigma^n} \int_0^{\infty} \!\D u\: \Gamma\left(p,\frac{u}{(2+\theta)^2} \right) u^{n-p}
\nonumber \\
&= \frac{1}{\Gamma(p)} \left( \frac{\sigma}{(2+\theta)^{2}}\right)^{p-1} \sum_{n=0}^{\infty} 
\frac{(-1)^n}{n!} \frac{\Gamma\bigl( 1+n \bigr)}{n+1-p} \left( \frac{\sigma}{(2+\theta)^{2}} \right)^{-n}
\label{gl:D.5b} \\
&= \frac{1}{\Gamma(p)} \left( \frac{\sigma}{(2+\theta)^{2}}\right)^{p-1} \frac{\Gamma(1-p)}{\Gamma(2-p)} {}_2F_1\left( 1-p,1;2-p;-\frac{(2+\theta)^2}{\sigma} \right)
\label{gl:D.5c}
\end{align}
\end{subequations}
where we used (\ref{gl:D.1d}) to arrive at the explicit form (\ref{gl:D.5a}). Afterwards, the exponential is expanded, the explicit form of the single-time correlator $F(1;u)$ is used 
and term-wise integration leads to a convergent series which can be recognised as hypergeometric function \cite{Abra65}. 

We give the end result, to be used in section~\ref{sec:4}, 
either as the integral representation (\ref{gl:D.5a}), a convergent power series (\ref{gl:D.5b}) for $\sigma\to\infty$ and finally recognise it as a
hypergeometric function in (\ref{gl:D.5c}). 

\noindent
{\bf 3.} For $d_s>2$, the r\^oles of $\mathscr{F}_0$ and $\mathscr{F}_1$ are exchanged and we can take over the last calculation to read off 
$\lap{\bigl(\mathscr{F}_1(\kappa)\bigr)}(\sigma)$, but with a different interpretation of $p$. 

\end{document}